\documentclass[pre,twocolumn,showpacs]{revtex4}

\usepackage{epsfig,amsmath,amssymb,graphics,color,calc,dcolumn,bm}

\newcommand{\be}{\begin{equation}}
\newcommand{\ee}{\end{equation}}
\newcommand{\ba}{\begin{eqnarray}}
\newcommand{\ea}{\end{eqnarray}}
\newcommand{\ban}{\begin{eqnarray*}}
\newcommand{\ean}{\end{eqnarray*}}

\begin{document}

\title{Effective interaction potential for amorphous silica from ab initio 
simulations}

\author{Antoine Carr\'e$^{1}$, Simona Ispas$^{2}$, J\"urgen Horbach$^{3}$, and Walter Kob$^{2}$}
 \email{walter.kob@umontpellier.fr}
\affiliation{$^{(1)}$Schott AG, Hattenbergstrasse 10, 55122 Mainz, Germany\\
             $^{(2)}$Laboratoire Charles Coulomb, Universit{\'e} Montpellier and CNRS, 
              UMR 5221, \\ 34095 Montpellier, France\\        
            $^{(3)}$Institut f\"ur Theoretische Physik II: Weiche Materie, Heinrich Heine-Universit\"at
                     D\"usseldorf, Universit\"atsstra\ss e 1, 40225 D\"usseldorf, Germany\\
}

\begin{abstract}
We discuss a novel approach that allows to obtain effective potentials
from {\it ab initio} trajectories. Our method consists in fitting the
weighted radial distribution functions obtained from the {\it ab initio}
data with the ones obtained from simulations with the effective potential,
and using the parameters of the latter as fitting variables.  As a case
study, we consider the example of amorphous silica, a material that is
highly relevant in the field of glass science as well as in geology.
Our approach is able to obtain an effective potential that gives
a better description with respect to structural and thermodynamic properties
than the potential proposed by van Beest, Kramer, and van Santen,
and that has been very frequently used as a model for amorphous silica. In
parallel, we have also used the so-called ``force matching'' approach
proposed by Ercolessi and Adams to obtain an effective potential. We
demonstrate that for the case of silica this method does not yield
a reliable potential and discuss the likely origin for this failure.
\end{abstract}


\maketitle

\section{Introduction}

In a classical molecular dynamics (MD) computer simulation, Newton's
equations of motion are solved for a many-particle system.  Nowadays, such
MD simulations allow to consider systems that contain generally between
10$^3$ and several million particles~\cite{Frenkel,binder04}. Depending
on the number of particles, typical time scales covered by the
simulation range between a nano-second and several micro-seconds. In a
classical simulation the interactions between the particles are given
by an effective potential function that does of course not take into
account in an explicit manner the electronic degrees of freedom. For
simplicity and efficiency of the simulations, one often considers
pairwise-additive interactions that depend only on the distance between
pairs of particles. In fact, if one aims to reveal universal properties
of, e.g., supercooled melts or glasses \cite{kob03}, or if the goal is
to check analytical theories that are based on simple interaction models,
such pairwise-additive interaction potentials are the most useful choice.

Although experience shows that even simple pair potentials can provide
a surprisingly realistic description of rather complex materials,
there is at present no systematic approach that allows to obtain an
effective two-body potential for a given material.  An example for
this is amorphous silica that forms a disordered tetrahedral network
structure in which tetrahedral SiO$_4$ units are connected to each
other {\it via} the oxygen atoms at the corner of each tetrahedron
\cite{silicabook,glassbook}. Although strong directional covalent
Si-O bonds are essential for the formation of the tetrahedral network
structure, pair-interaction models such as the one proposed by van
Beest, Kramer, and van Santen (BKS) \cite{bks} do reproduce many
static and dynamic properties of real silica (see, {\it e.g.},
Refs.~\cite{vollmayr96,hemmati97,horbach99,horbach08,carre08,saika11,tokuyama11,guillot12}).

Due to its importance for glass science as well as technology,
intensive efforts have been undertaken in the past to investigate
silica by means of classical MD simulations.  To this end, different
effective interaction models have been proposed in an {\it ad hoc}
manner and also a variety of fitting schemes have been proposed,
aiming at a more systematic way to obtain effective potentials
\cite{bks,hemmati97,carre08,woodcock76,soules79,soules11,sanders84,matsui85,matsui87,lasaga87,catlow88,feuston88,tsuneyuki88,chelikowsky90,vashishta90,sixtrude91,kramer91,schroder92,boer94,boer95,wilson93,wilson96,blonski97,duffrene98,demiralp99,tangney02,flikkema03,huang04,hoang06,pedone06,paramore08,coslovich09}.
The first effective potentials were mainly geared towards the description
of the crystalline phases of silica, such as $\alpha$-quartz and
$\beta$-cristobalite and thus in some cases the functional Ansatz for
these potentials were specifically designed to reproduce the symmetric
tetrahedral environment found in crystals. The parameters of these
potentials were adjusted to reproduce some macroscopic experimental data
such as elastic constants, or microscopic data such as lattice parameters.
Consequently these potentials reproduce by design such features, but
give in fact very little considerations to the microscopic interactions while,
ideally, an effective potential should describe at best the microscopic
interactions and should consequently be able to describe and predict
reliably macroscopic observables.

A significantly improved methodology has been proposed by
Tsuneyuki \textit{et al.} who derived a pair potential based on
quantum calculations of the vibrational modes of a single Si(OH)$_4$
unit~\cite{tsuneyuki88}. Later, this potential was improved by BKS
\cite{bks} using a similar approach but these authors included also
information on bulk data. Although the BKS potential has been found
to be surprisingly reliable for the description of many properties
of amorphous silica (see above), it has also shortcomings in that,
{\it e.g.}~the equation of state \cite{carre08} and the vibrational
density of state \cite{Benoit02} show relatively strong deviations
from experimental data. That the fitting scheme, originally proposed by
Tsuneyuki {\it et al.}, leads to a good description of silica, is also
somewhat accidental, since in general one cannot expect that a quantum
calculations of a single tetrahedral unit allows to obtain an effective
interaction potential of a bulk material. 
Although
a similar methodology has also been successfully applied to GeO$_2$
\cite{oeffner98,hawlitzky08}, it is not obvious that it
can be generalized to other
oxides such as Al$_2$O$_3$ and B$_2$O$_3$ or more complex systems such
as borosilicates, alkali silicates, phosphates and fluorophosphate
glasses which are of prime technological importance.

It is evident that the idea to use {\it ab initio} data to obtain
effective classical potentials is very appealing since {\it ab initio}
simulations allow, at least in principle, to obtain the properties
of any material. In such simulations one describes the ion cores as
classical particles and the electronic degrees of freedom are treated
quantum mechanically within the framework of a density functional
theory~\cite{marx09}. {\it Ab initio} simulations are nowadays routinely
done for systems of 100 to 300 atoms on a time scale of several tens of
picoseconds~\cite{kalikka_14,bouzid_15,pedesseau_15ab}.  They allow to
study systems in a well-defined thermodynamic ensemble, with the full
information of the trajectories of all the ions in the system. From
this information it is thus in principle possible to obtain effective
potentials. In practice, however, it is not really feasible to integrate
out the electronic degrees of freedom and thus to determine an exact
effective potential function from the \textit{ab initio} trajectories. Hence, one
usually makes an Ansatz for an effective potential that contains a number
of free parameters.  Then these parameters are adjusted by a fitting
procedure such that the optimal agreement between the {\it ab initio}
data and classical MD calculations is obtained.

The Ansatz one chooses for the effective potential will depend on the
nature of the system under consideration and often this choice is based
on physical intuition. As mentioned above, systems such as silica can be
described quite well by two-body potentials, such as the aforementioned
BKS model for silica which consists of long-ranged Coulomb and short
ranged Buckingham terms \cite{buckingham38}, with the latter describing
the attractive and repulsive interactions between the ion cores on
length scales of~2-8~{\AA}. Once one has decided on the functional form
of the potential one has to address the problem on how to fit the free
parameters of this function to obtain an optimal reproduction of the {\it
ab initio} data. This is of course a highly non-trivial task because
one is confronted with a high-dimensional fitting problem since the
potential of even the simplest systems will have on the order of 10-20
parameters that have to be determined.  One must expect that the cost
function that one optimizes (see below) will have many local minima in
parameter space and thus there is the risk to be trapped in one of these
and hence to not find the optimal set of parameters. In addition, one
has to keep in mind that the Ansatz in terms of two-body or even three-
or four-body interactions neglects higher-order many-body interactions,
required to exactly describe the effective interactions between the ion
cores. As a consequence, the cost function has to be chosen such that
the {\it relevant physical properties} of the system are reproduced in
simulations with the resulting effective potential. Thus, the choice of
the cost function and its minimization are crucial to parameterise the
potential such that it leads to a reliable description of the system
under consideration.

In the present work, we compare two different approaches to determine
the parameters of effective pair potentials. The system considered
is amorphous silica (SiO$_2$) and the {\it ab initio} data has been
obtained from a Car-Parrinello molecular dynamics (CPMD) simulation
\cite{car85,cpmd}. (In the following we refer to these simulations as
CPMD simulations). As an Ansatz for the functional form of the three
pair potentials that describe the O-O, Si-Si, and Si-O interactions,
we use a combination of Coulomb and Buckingham terms, i.e.~the same
functional form as in the BKS model. This choice for the form of the
potential is motivated by the fact that the BKS potential does give a
good description of many properties of silica and hence it can serve as a
reference on what can be achieved by this simple functional form. We do,
however, of course not exclude the possibility that other type of pair
potentials will give an even better description of this material.

The first approach we will use for finding the optimal parameters is the
so-called ``force-matching method'', proposed by Ercolessi and Adams
\cite{ercolessi94}. In that approach, the parameters of the effective
potential are optimized such that the resulting force on each particle
matches as closely as possible the one obtained from the {\it ab initio}
simulation. Below we will show that if one uses the parameters of the
BKS model as start parameters for the fitting, one does indeed find an
effective potential that gives a very good description of these forces.
However, we show that the structure resulting from the force matching
procedure gives only a very bad description of silica and is in
fact much worse than the BKS model. This becomes, e.g., evident if one
uses this potential to calculate structural quantities such as the pair
correlation or angular distribution functions, and compares them with
those from the CPMD simulation.

The second approach that we consider for optimizing the parameters
uses as input the structural data from the {\it ab initio} simulations,
notably the pair correlation function. This method has been applied in
the past to obtain a reliable potential for silica~\cite{carre08}.  Here,
we extend this work and clarify why the structure matching method
leads to a much better model than the force-matching method, at least in
the case of silica. Thus, we demonstrate that the structure-matching
method provides a systematic scheme to obtain effective potentials from
{\it ab initio} trajectories.

The outline of this paper is as follows: In Sec. \ref{AIS}A we present the
methodology as well as the simulations parameters we use for the {\it ab
initio} simulations. The results of these simulations are used as input
to the force matching procedure described in Sec.~\ref{FMP}. The fitting
procedure based on the structural data is described in Sec.~\ref{SFP}. The
structural and dynamical features given by the two new potentials 
are presented in Sec. \ref{sec3} and in Sec. \ref{CAO} we give a summary.

\section{Obtaining the potential\label{AIS}}

In this section, we will first give the details on the {\it ab initio}
simulations that have been used to generate the data that is afterwards
used to make the fits. Subsequently, we will present two fitting procedures
that allow to obtain an effective potential.

\subsection{\textit{Ab initio} Simulation Details\label{Sim_Det}}

The system considered in the \textit{ab initio} simulations contains
$N=114$ particles, {\it i.e.}~38 SiO$_2$ units, confined in a cubic box with
periodic boundary conditions. The size of the box is $L=11.982$\,{\AA}
which corresponds to a mass density of $2.2$\,g/cm$^3$, the experimental
value of the density of silica at room temperature~\cite{CRC}. To obtain
an initial configuration, we have first carried out a classical simulation
at 3600\,K using the BKS potential~\cite{bks}. From this simulation we have extracted
an equilibrium configuration and used it as starting point for the
Car-Parrinello molecular dynamics (CPMD) runs \cite{car85}.
Although the BKS potential is not perfectly reliable regarding the
structural properties of SiO$_2$, it does give a reasonable description
of the structure and hence the {\it ab initio} simulation do not have
to be very long in order to reach an equilibrium state.

Within the {\it ab initio} simulations, we have described the electronic
structure of the system by means of the Kohn-Sham formulation of
the density functional theory using the local density approximation
\cite{Martin-book}. The Kohn-Sham orbitals were expanded in a plane-wave
basis at the $\Gamma$ point of the supercell. Core electrons were not
treated explicitly but were replaced by atomic pseudopotentials of
the Bachelet-Hamann-Schl\"{u}ter type for silicon  \cite{Bachelet}
and the Troullier-Martins type for oxygen  \cite{Troullier}. This
choice of pseudopotentials and exchange and correlation functionals are
motivated by previous {\it ab initio} simulations of amorphous SiO$_2$
\cite{Benoit02,Benoit_Ispas_1,Benoit_Ispas_2}. The simulations have
been carried out with version 3.9.2 of the CPMD code~\cite{cpmd},
using a time step of $0.0725$\,fs and a fictitious electronic mass
of $\mu=600$\,atomic units. The cutoff radius used in the CPMD
simulation for defining the spatial extent of the plane waves in the
reciprocal space was $70$\,Ry.  The simulations were done in the $NVT$
ensemble at 3600\,K and the temperature was controlled by means of
a Nos\'e-Hoover chain \cite{Nose_Hoover_1,Nose_Hoover_2} coupled to
each nuclear degree of freedom of the system~\cite{Martyna1992}. As
documented in Refs.~\cite{Benoit_Ispas_2,Poehlmann}, this procedure
shortens considerably the time for equilibration. As already described
in previous studies~\cite{Sarnthein_Pasquarello}, one finds that in
silica the electronic gap can close. To counterbalance the energy flow
from the ions to the electrons \cite{Sprik}, we have therefore connected
also the electronic degrees of freedom to a Nos\'e-Hoover chain. The
sample was equilibrated within the CPMD for 3.5\,ps, a time span that
for this temperature is sufficiently long to allow all the species
to reach a diffusive behavior. The equilibration was followed by a
production run of 16\,ps and this trajectory was used to calculate the
observables of interest for the system discussed in more detail below
(radial distribution functions, forces, {\it etc.}). In practice, we have
extracted from this trajectory one hundred configurations, \textit{i.e.}~one 
configuration every $160$\,fs, and the wave-function of each of
these configurations has been quenched to the Born-Oppenheimer surface.
Although a $70$\,Ry energy cutoff is sufficient to obtain converged
values for the forces, we have used a cutoff of $130$\,Ry to ensure the
proper convergence of the diagonal elements of the stress tensor. More
details on this can be found in Ref.~\cite{Carre_Phd}.

\subsection{Potential from force matching \label{FMP}}

The so-called ``force matching (FM) procedure'' proposed by Ercolessi
and Adams, Ref.~\cite{ercolessi94}, starts from the physical idea that
the forces on particle $i$ as obtained from the effective potential,
$\mathbf{F}_{i}^{\rm{FM}}$, should be able to reproduce as closely as possible
the real force on the particle, \textit{i.e.}~in the present case the 
force obtained from the {\it ab initio} simulations, $\mathbf{F}_{i}^{\rm{CP}}$.

To achieve this goal we introduce the cost function $\chi_F^2$ which
is defined as
\begin{equation}
  \chi_F^2 =  \frac{1}{N} \sum_{i=1}^N 
    \frac{\langle|\mathbf{F}_{i}^{\rm FM} -  \mathbf{F}_{i}^{\rm CP}|^2 \rangle}{\sigma_i^2} \quad .
\label{Chisq_Definition_Force_Fit}
\end{equation}
%
Here $\langle .\rangle$ stands for the canonical average, which in our
case is approximated by the average over the 100 selected configurations
mentioned above. In contrast to the original method described in
Ref.~\cite{ercolessi94}, the force difference for particle $i$ is here
divided by $\sigma_i$, the average standard deviation of the force
acting on particle $i$. This modification is motivated by the fact that
different type of atoms will experience different {\it average} forces
and fluctuations, and hence the cost function, should take this into
account. In practice, we have considered that all atoms of a given specie have the same $\sigma_i$, and have used  the following definition for $\sigma_\alpha$, for $\alpha=$~Si, O:
\begin{equation}
\sigma_\alpha=\sqrt{ \langle |\mathbf F_\alpha^{\rm CP}|^2 \rangle
  - |\langle \mathbf F_\alpha^{\rm CP} \rangle|^2   } = \sqrt{ |\langle 
\mathbf F_\alpha^{\rm CP}\rangle|^2}
\quad ,
\end{equation}
where the second equality holds because  $\langle  {\mathbf F}_{i}^{\rm{CP}}
\rangle=0$. The numerical values determined for $\sigma_{\rm Si}$
and $\sigma_{\rm O}$ are 3.55\,eV/\AA\, and 2.69\,eV/\AA, respectively.

The functional form of the effective potential considered for
the fit is basically the same as the one proposed by van Beest, Kramer,
and van Santen~\cite{bks} and is given by:
\begin{equation}
V^{\rm eff}  = 
    \sum_{i<j} \frac{q_i q_j}{4 \pi \varepsilon_0 r_{ij}} + 
                A_{ij} \exp\left( -B_{ij} r_{ij}\right) - 
               \frac{C_{ij}}{r_{ij}^6} + 
               \frac{D_{ij}}{r_{ij}^{24}}
\label{BPOTENTIAL}.
\end{equation}
The short range (SR) part of the potential (i.e.~the second and third
term) is of the Buckingham type and has been truncated and shifted at a
distance $r_{\rm cut }=6.5$\,\AA. In order to have a smooth function, we
have then multiplied this part by a function $G(r)$ such that the resulting
function and its derivatives go smoothly to zero at $r \le r_{\rm
cut }$ and are zero for $r>r_{\rm cut}$. Finally, our smoothed
SR potential is given by
\begin{eqnarray}
V_{\rm SSR}(r) & = & \left\{
    \begin{array}{r@{\quad:\quad}l}
      [V_{\rm SR}(r) - V_{\rm SR}(r_{\mbox{\scriptsize{cut}}})]G(r) & r \le
      r_{\mbox{\scriptsize{cut}}}\\
      0 & r > r_{\mbox{\scriptsize{cut}}}
    \end{array}
  \right.\\
{\rm with} \; G(r) & = &\exp \left(
    -\frac{\gamma^2}{\left(r-r_{\mbox{\scriptsize{cut}}}\right)^2}\right).
\end{eqnarray}
All the simulation results presented in this paper have been done with
a value of $\gamma^2 =0.05$ \AA$^2$. The (strongly repulsive) last term
$D_{ij}/r^{24}_{ij}$ in Eq.~(\ref{BPOTENTIAL}) has been introduced
to prevent that two particles fuse together due to the van der Waals
forces. The values we have used for the parameters $D_{ij}$ are (in
eV.{\AA}$^{24}$) given by 113, 29, and 3423200 for the O-O, Si-O, and
Si-Si pairs, respectively. Note that none of the subsequent results will depend in a
relevant manner on the choice of these parameters.  Finally, we mention
that the Coulombic term in Eq.~(\ref{BPOTENTIAL}) has been calculated
using the Ewald summation method  \cite{Frenkel}.  For convenience we will
use in the following the notation $\boldsymbol{\xi}=\{\xi_1,...,\xi_{10}\}$
to designate the 10 parameters of the effective potential that have to be
minimized (\textit{i.e.}~$q_{\rm Si}$, $A_{ij}$, $B_{ij}$, and $C_{ij}$).

The cost function $\chi_F^2$ given by
Eq.~(\ref{Chisq_Definition_Force_Fit}) has been minimized using as
starting point the BKS parameters~\cite{bks}.  The optimization
has been performed using the Levenberg-Marquardt algorithm
\cite{Numerical_Recipes}, \textit{i.e.}~a minimization approach that
combines the Gauss-Newton algorithm and the steepest descent method. The
weighting between these two algorithms is regulated by a parameter
which is iteratively being updated~\cite{Numerical_Recipes}. Far from
any local minima the Levenberg-Marquardt behaves similar to the steepest
descent method, close to a local minimum, where the gradient becomes very small,
the minimization is done by means of the more computational demanding
Gauss-Newton algorithm. Note that this method requires the calculation
of the first derivatives of the cost function $\chi^2_F$ with respect
to the parameters $\boldsymbol{\xi}$. Due to the very simple form of
the Buckingham-Coulomb Ansatz it is possible to calculate analytically
these expressions so as to gain precision and computing efficiency. These
analytical forms have been thereafter numerically evaluated for the
purpose of the fit.

In the following we will refer to the potential obtained with
this approach as the ``FM potential''. Its parameters are given
in Table~\ref{RGR_PARAM} and its properties will be discussed in
Sec.~\ref{sec3}.

\subsection{Potential from the structure \label{SFP}}

In the second approach considered to obtain an effective potential no
attempts have been made anymore in order to directly optimize the forces
working on the particles. Instead the attention has been focused on the
description of the local structural environment. It is evident that there
is a large choice regarding the type of structural observables that can
be considered for such an optimization. In the following, we will focus
on two-point correlation functions, a choice that is certainly reasonable
and not specific to the open network structure of SiO$_2$.

The definition of the cost function $\chi^2_n$ used in our structural fitting 
procedure is given by
\begin{equation}
  \chi^2_n =  \sum_{\alpha,\beta}  \int_{0}^{L/2} 
    \left[ r^{n} ( g_{\alpha  \beta}^{\rm{CP}}(r)- 
          g_{\alpha \beta}(r;\boldsymbol{\xi})) \right]^2dr \quad .
\label{chi_def}
\end{equation}
Here $g_{\alpha\beta}(r;\boldsymbol{\xi})$ are the partial pair correlation functions
defined as~\cite{hansen_XX}
\begin{equation}
g_{\alpha \beta}(r; \boldsymbol{\xi}) = 
\frac{L^3}{N_\alpha (N_\beta - \delta_{\alpha\beta})}\frac{1}{4\pi r^2}
\sum_{i=1}^{N_\alpha} \sum_{j=1}^{N_\beta}
\left \langle \delta(r- |\vec r_i -\vec r_j|) \right \rangle \quad ,
\label{eq4}
\end{equation}
where $L$ is the size of the simulation box, $N_\alpha$ is the number of
particles of species $\alpha$, and $\delta_{\alpha\beta}$ is the Kronecker
delta function. Note that the right hand side of Eq.~(\ref{eq4}) depends
on $\boldsymbol{\xi}$ via the thermal average. The exponent $n$ allows to choose how
the different distances are weighted: $n=0$ makes that the difference
on the right hand side of Eq.~(\ref{chi_def}) is just the difference in
the radial distribution function, \textit{i.e.}~a significant weight is
put on the first nearest neighbor distance, whereas with increasing $n$
larger distances are given a larger weight. The choice of $n$ will be
discussed in more detail below.

\begin{figure}
\centering
\includegraphics*[width=0.43\textwidth]{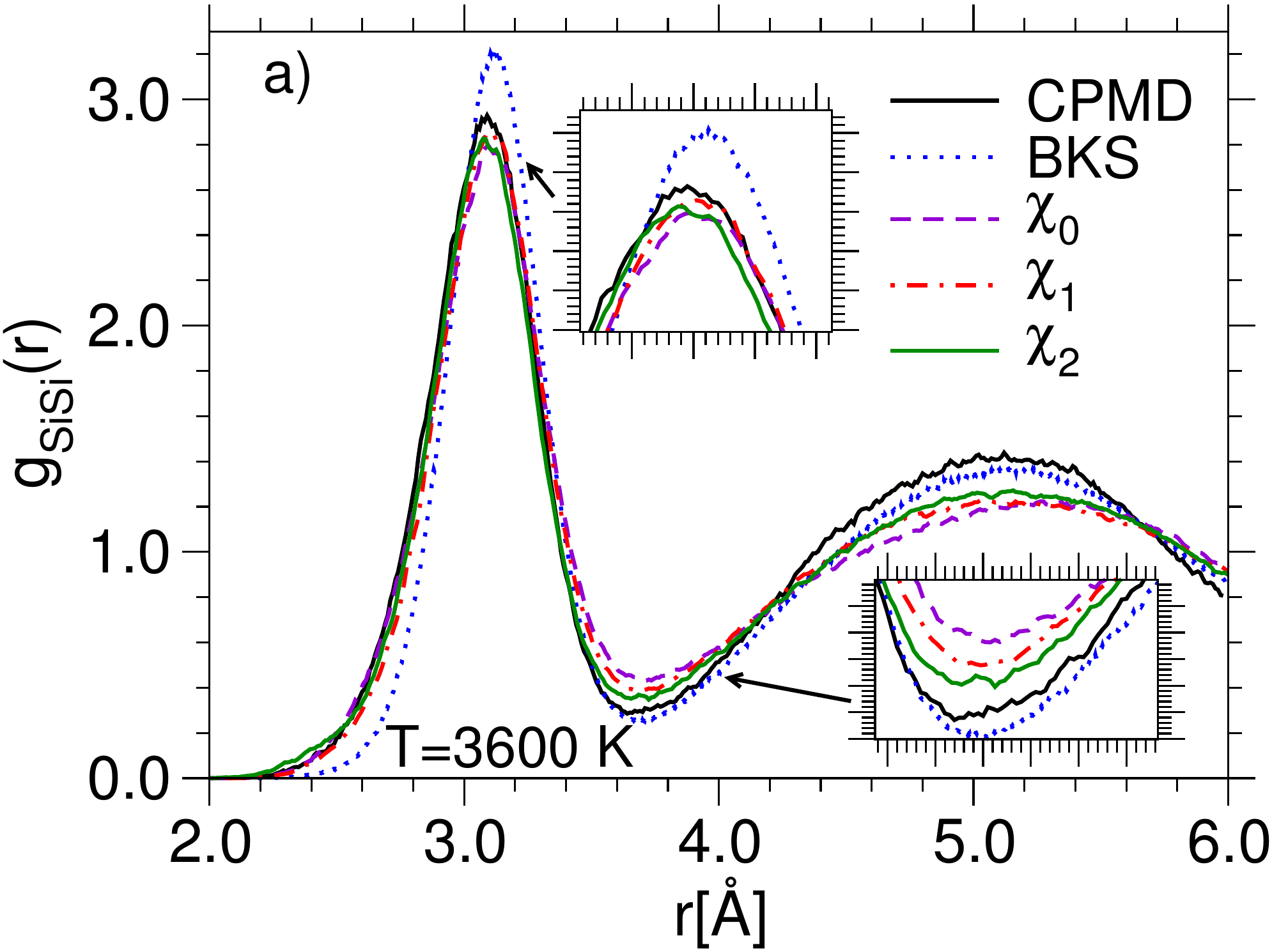}
\includegraphics*[width=0.43\textwidth]{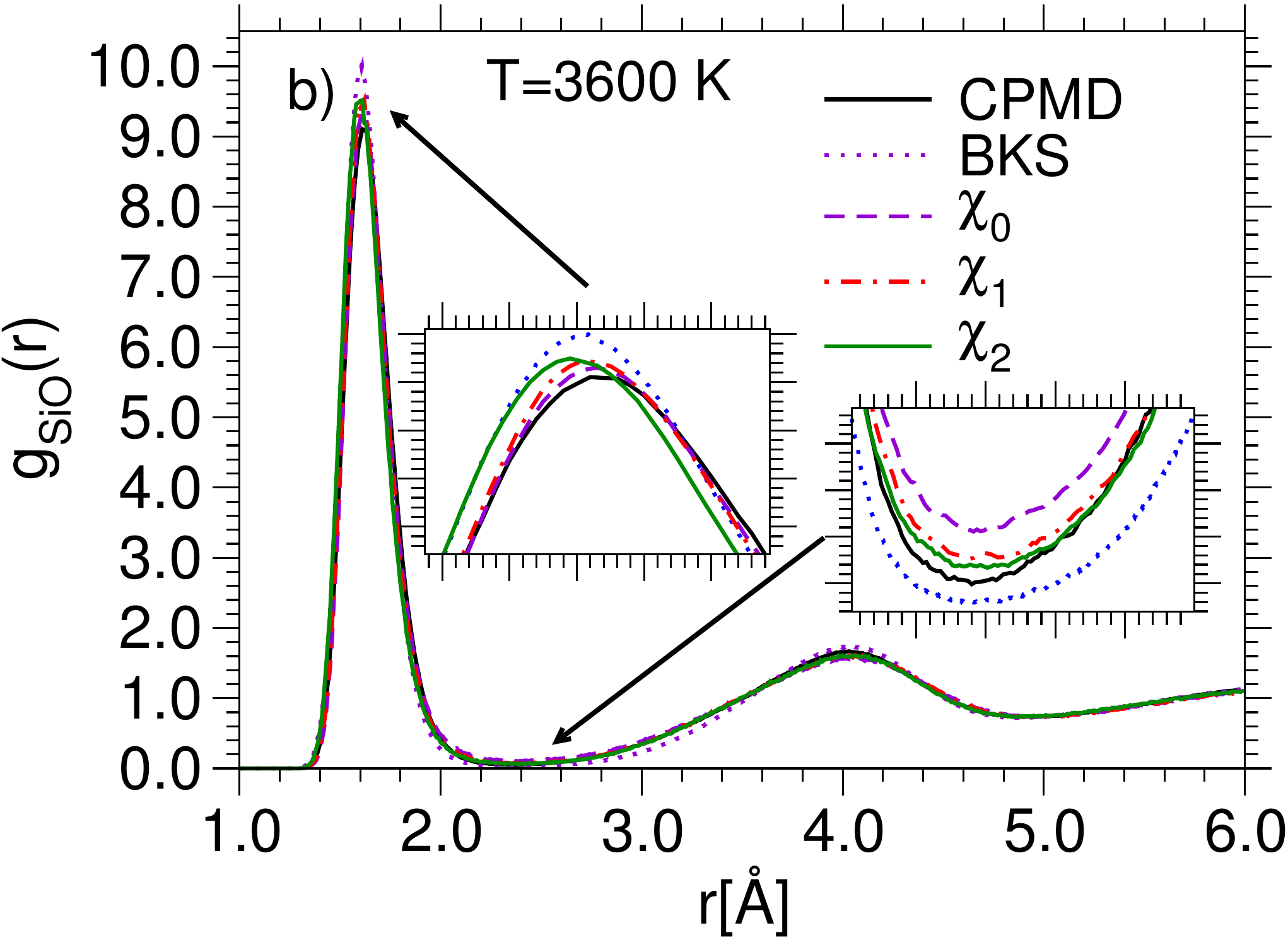}
\includegraphics*[width=0.43\textwidth]{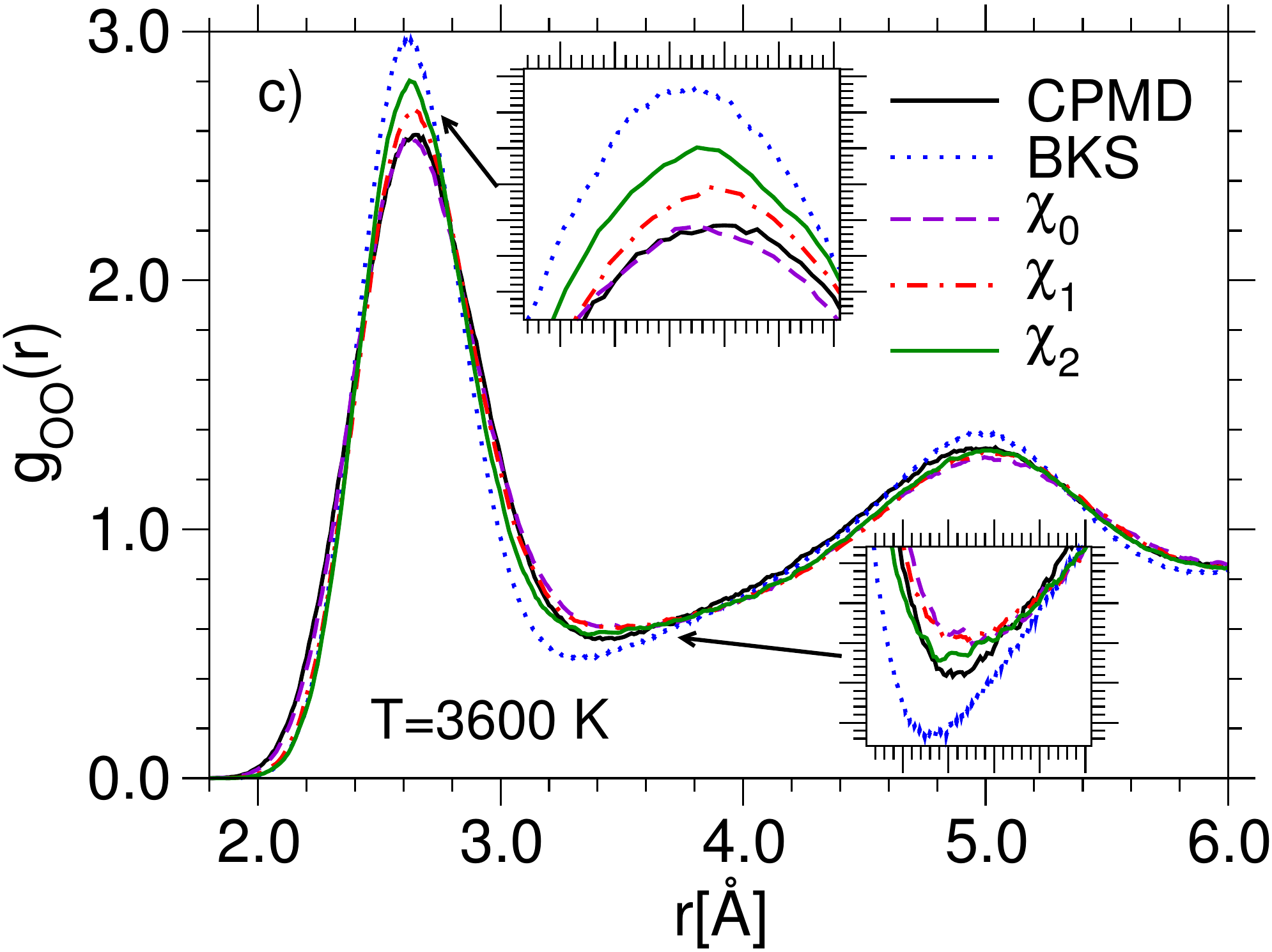}
\caption{Partial pair distribution functions at $T=3600$\,K computed
using different sets of parameters (BKS, $\chi_n, n \in \{0,1,2\}$) and 
compared to the CPMD results.
\label{gr_criterions}}
\end{figure}

To calculate the cost function $\chi_n^2$ from Eq.~(\ref{chi_def}),
we need first to determine the functions $g_{\alpha \beta}(r;
\boldsymbol{\xi})$. For this we have made an $NVT$ simulation
at $T=3600$\,K, using $N=1152$ atoms confined in a cubic box of
$L=25.904$\,{\AA},  \textit{i.e.}~$\rho=2.2$\,g/cm$^3$.  These
simulations, with a time step of 1\,fs, had a length of 10\,ps. The
first 20\% of these runs were discarded since they might be affected by
out-of-equilibrium effects (the starting configuration used for these MD
runs corresponds to a SiO$_2$ melt equilibrated using another empirical
potential, and hence the change in the effective potential will lead
to an out-of-equilibrium situation.)

The search for the optimal set of parameters $\boldsymbol{\xi}$ has again been
done by means of the Levenberg-Marquardt algorithm. Since this requires
the derivative of $\chi_n^2$ with respect to the various $\xi_i$,
and hence the variation of $g_{\alpha \beta}(r,\boldsymbol{\xi})$
with $\boldsymbol{\xi}$, we have determined the latter from a finite
difference scheme:
\begin{equation}
  \frac{\partial g_{\alpha \beta}(r;\boldsymbol{\xi)}}{\partial \xi_i} \approx 
  \frac{g_{\alpha \beta}(r;\xi_i+\varepsilon_i)-
        g_{\alpha\beta}(r;\xi_i-\varepsilon_i)}{2 \varepsilon_i}.
\label{eq6}
\end{equation}
Here the variations $\varepsilon_i$ have to be chosen with care for two
reasons: On the one hand, these variations must be sufficiently small so that
the approximation of Eq.~(\ref{eq6}) remains accurate, {\it i.e.}~that
the finite difference is a good approximation of the derivative. On the
other hand, a variation of the fitting parameters that is too small will
make that the variation of $g_{\alpha\beta}(r,\boldsymbol{\xi})$ is very
small and hence the right hand side of Eq.~(\ref{eq6}), which is subject
to noise because of the finite length of the run, cannot be determined
with sufficient precision.  In practice, the choice of $\varepsilon_i$
will also depend on how the discretization of the distance $r$, needed to
calculate the integral~(\ref{eq4}), are made.  Indeed, a small spatial
discretization grid will give rise to a larger statistical uncertainty
per bin of the $g(r)$, whereas a bin size that is too large will give
an inaccurate evaluation of the integral. To calculate the integral in
Eq.~(\ref{chi_def}), we have divided the distance $r$ in bins of width
0.02\,\AA. Given the length of the MD trajectory, it has been determined
that a good choice for the $\varepsilon_i$'s is to take them on the
order of a few percent of $\xi_i$. In Table~\ref{RGR_PARAM} we have
included in the last column also the parameters that have been used here.
More details can be found in Ref.~\cite{Carre_Phd}.

At this point, the importance of the equilibration phase must be
emphasized for the scanning the parameter space. Each change of
$\boldsymbol{\xi}$ will make that the initial particle configuration
will no longer be representative of an equilibrium situation,
this means that the system first needs to be re-equilibrated.
This re-equilibration has been achieved by replacing every 50 time
steps the velocities of all the particles by ones that have been
drawn from a Maxwell-Boltzmann distribution \cite{Andersen}. This
massive thermostatting is necessary since the change of the potential
generates a quite strong out-of-equilibrium situation, and more gentle
thermostats, like, \textit{e.g.}~Nos\'e-Hoover, are not very efficient
in such situations.  Only once the system is equilibrated, one can start
measuring the structural quantities, such as the radial distribution
functions, that are required to calculate the cost function.

Finally, we comment on the choice of the exponent $n$ in
Eq.~(\ref{chi_def}). Using the fitting procedure just described we have
determined the optimal values for the parameters $\boldsymbol{\xi}$
by considering in the cost function of Eq.~(\ref{chi_def}) three
values of $n$: $n=0$, 1, and 2. With these three sets of optimized
parameters we have carried out standard MD simulations and determined
the partial pair distribution functions at $T=3600$\,K. These functions
are shown in Fig.~\ref{gr_criterions}. Also included in the figures
are the $g_{\alpha\beta}(r)$ as predicted, for the same temperature,
by the CPMD simulation and the BKS potential. A comparison of the
different curves shows that the BKS potential gives a fair description
of the CPMD data, but that it is definitively less accurate than the
prediction of the three other potentials, notably in the vicinity of
the main peaks. Furthermore, it can be observed that with increasing
$n$, the differences of the corresponding $g_{\alpha\beta}(r)$ with the
CPMD data increase at the main peak and decrease at the first minimum,
a trend that is understandable since a larger $n$ implies that the cost
function is more sensitive for larger $r$.

\begin{figure}
\centering
\includegraphics*[width=0.40\textwidth]{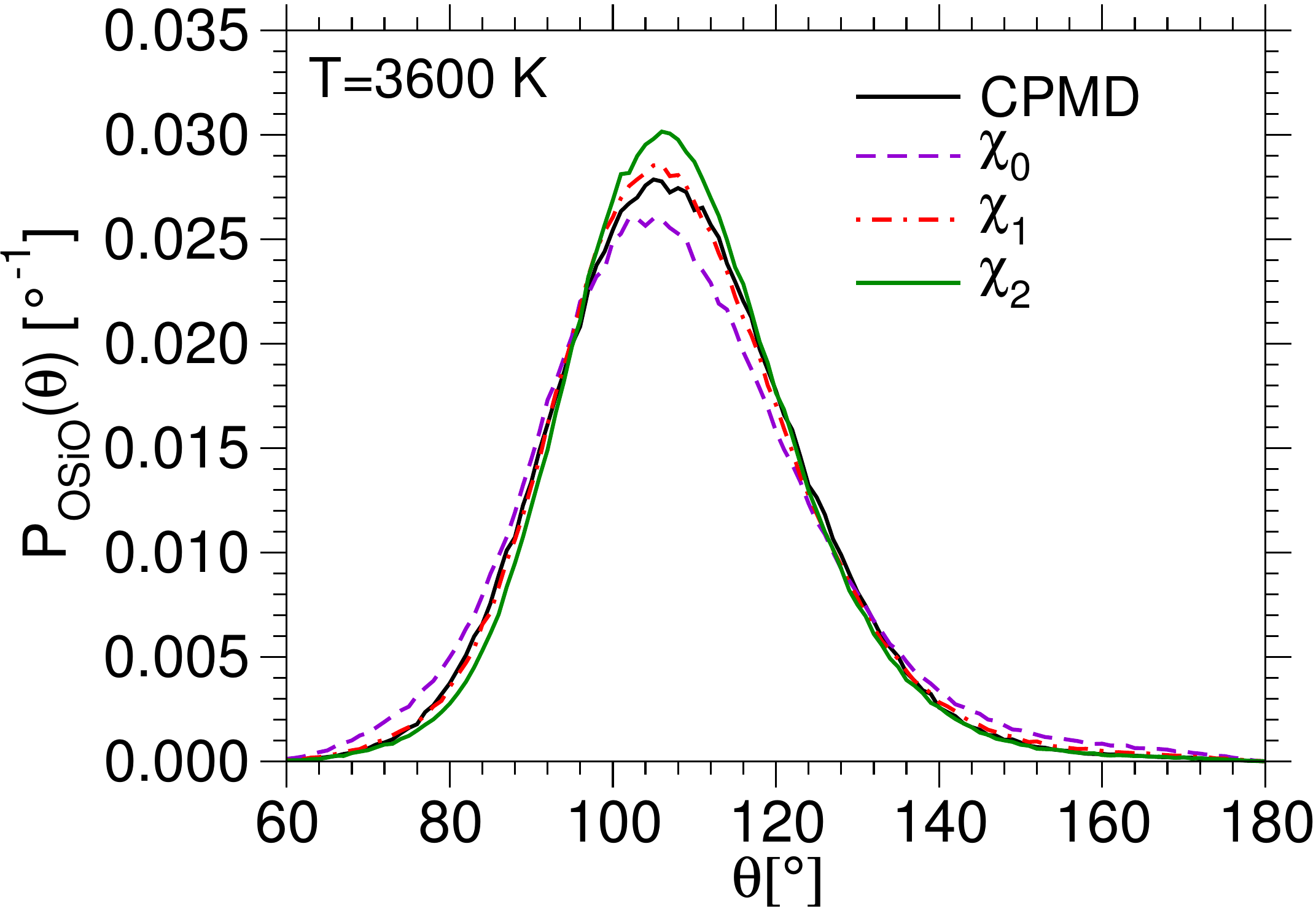}
\includegraphics*[width=0.40\textwidth]{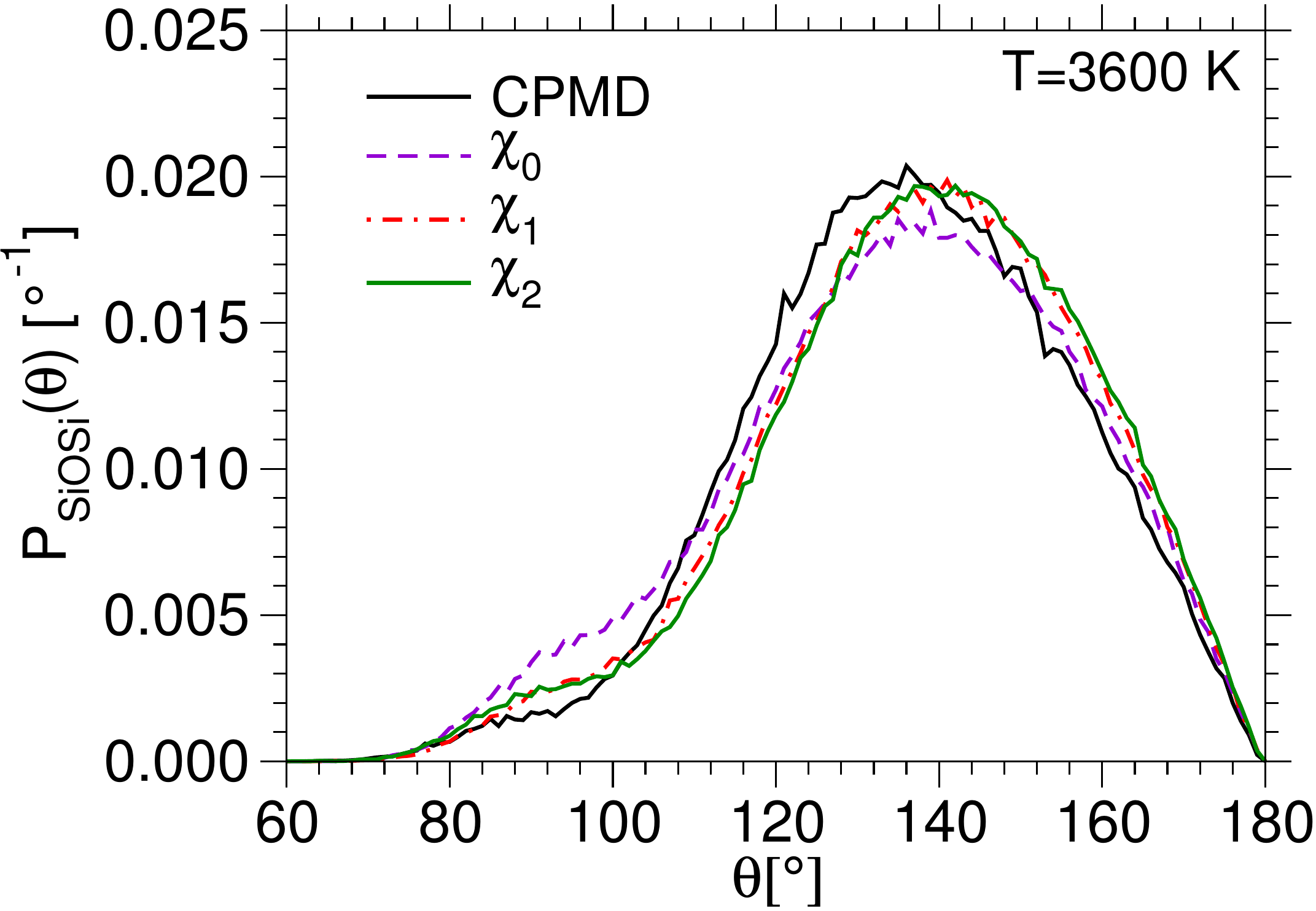}
\caption{Angular distribution functions for the $\widehat{\mbox{OSiO}}$
and $\widehat{\mbox{SiOSi}}$ angles at $T=3600$\,K as obtained from
different potentials.}
\label{angle_criterions}
\end{figure}

For probing the influence of $n$ on the structure it is also
useful to study the average angle spanned by two particles that
share a common neighbor. For the case of $\widehat{\mathrm{OSiO}}$
and $\widehat{\mathrm{SiOSi}}$, the corresponding distributions are
shown in Fig.~\ref{angle_criterions}. (The meaning of the various
peaks and shoulders will be discussed below but can also be found,
e.g., in Ref.~\cite{vollmayr96}.)  From this figure we recognize
that the distributions corresponding to the different values of
$n$ agree all quite well with the data from the CPMD simulations.
However, a closer inspection of the distributions, in particular the
one for $\widehat{\mathrm{OSiO}}$ around the peak, shows that $n=1$
gives a slightly better description of the CPMD distribution than
the other potentials. Since also the pair distribution function
for $n=1$ is very similar to the one from the CPMD simulations, see
Fig.~\ref{gr_criterions}, in the following part of the text the potential
derived using the $\chi^2_{1}$ criterion will be referred as being the
``CHIK potential''~\cite{carre08}. We note that data labelled  ``$\chi_{1}$" in  Figs.~\ref{gr_criterions}--\ref{angle_criterions} correspond to preliminary tests, and, due to different run lengths,  differences exist with respect to similar data labelled ``CHIK potential'' in the following sections (e.g. in Figs.~\ref{gr_CPMD_FF_BKS}--\ref{angle_CPMD_BKS_FF}).

\subsection{New effective potentials\label{nep}}

\begin{figure}
\centering
\includegraphics*[width=0.43\textwidth]{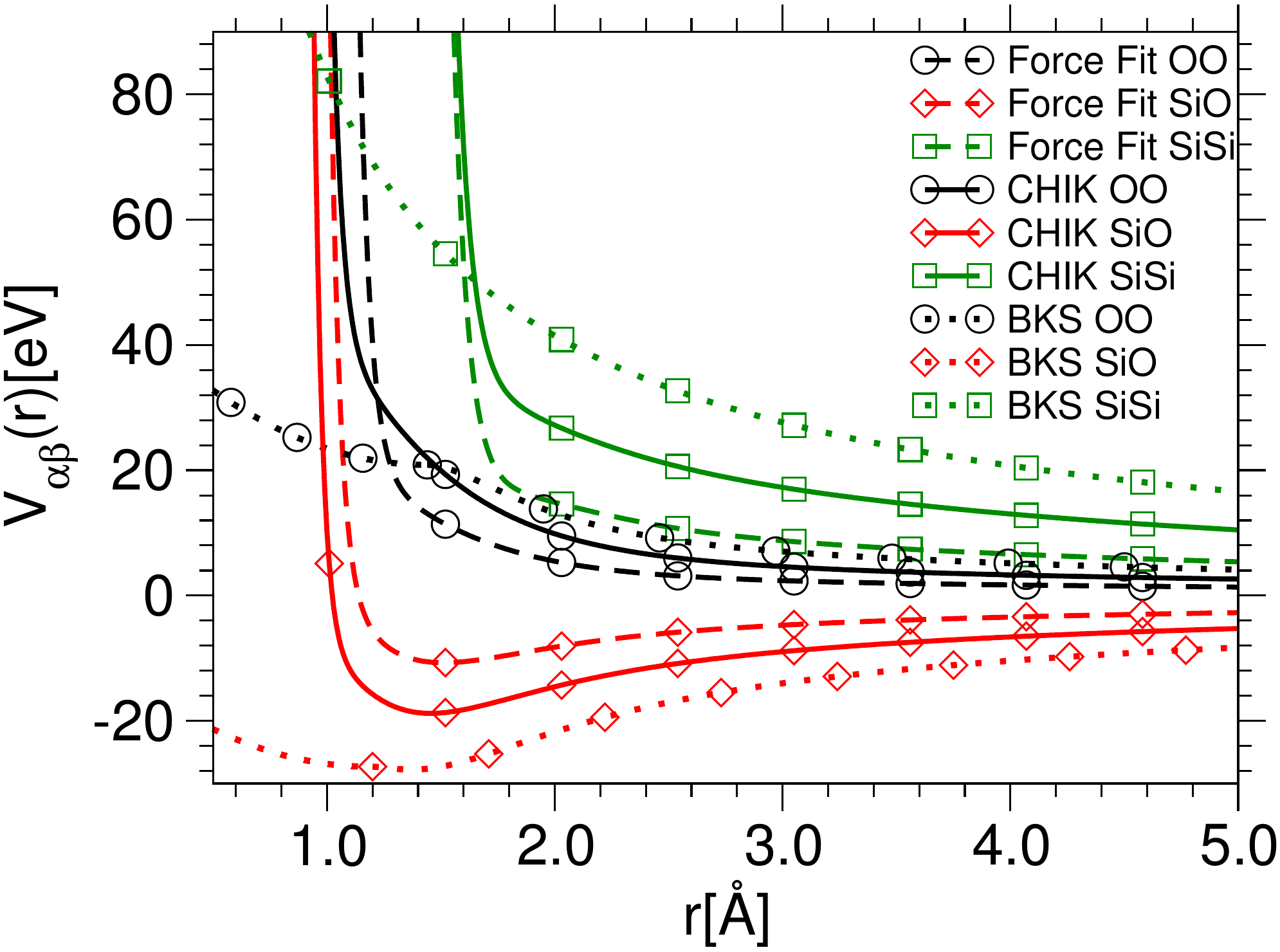}
\includegraphics*[width=0.43\textwidth]{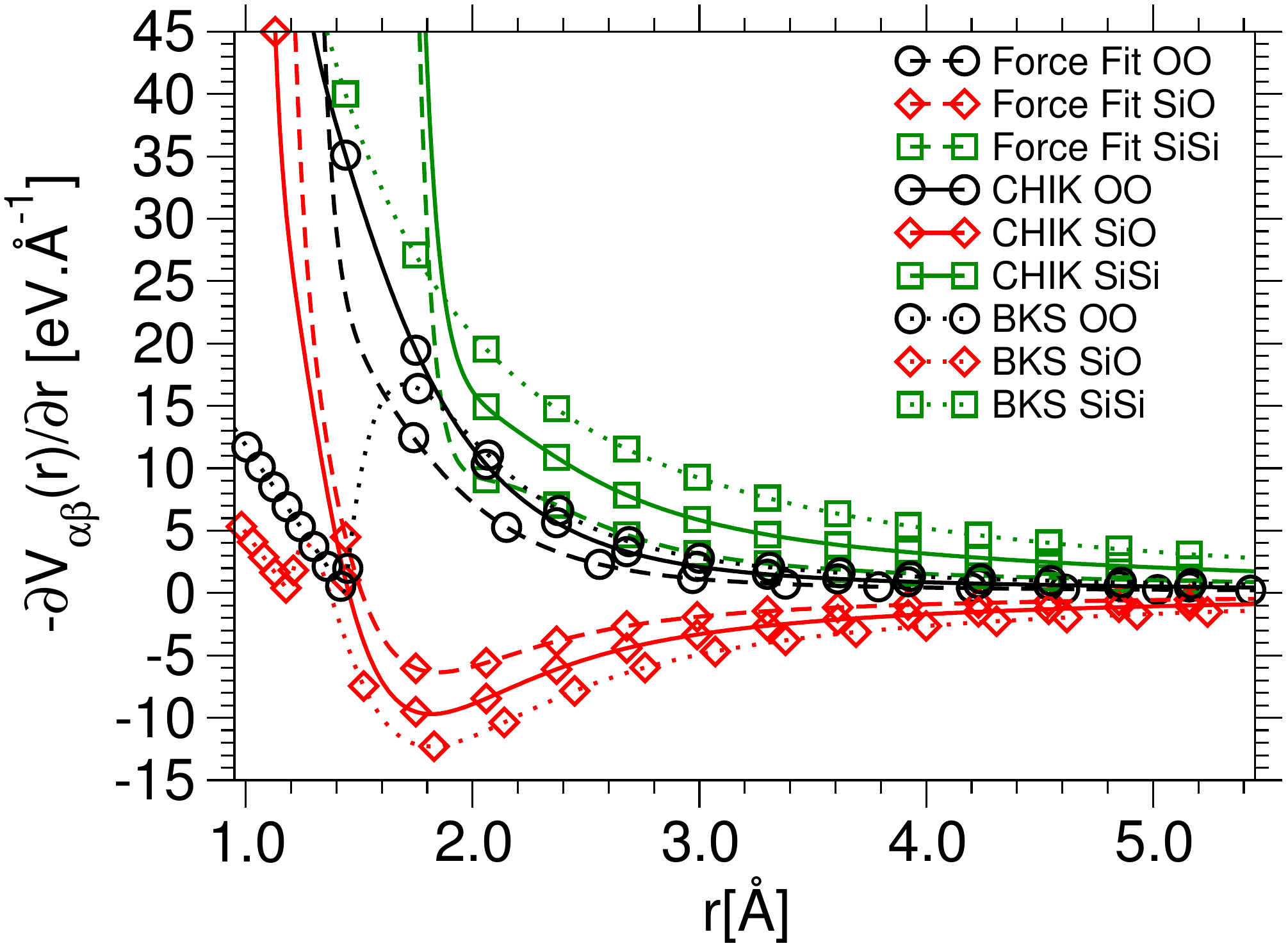}
\caption{a) The $r-$dependence of the various effective potentials considered 
in this work. b) The corresponding forces.}
\label{potential_energy_shape}
\end{figure}

The values for the optimal parameters $\boldsymbol{\xi}$ that we
have obtained from the force-matching approach derived following
Eq.~(\ref{Chisq_Definition_Force_Fit}), and the CHIK potential are
listed in Tab.~\ref{RGR_PARAM}. For the sake of comparison, we have also
included the values of the BKS potential from Ref.~\cite{bks}. In
order to understand better their similarities and differences,
the three potentials and the corresponding forces are plotted in
Fig.~\ref{potential_energy_shape}.

\begin{table}
  \begin{tabular}{|cl|c|c|c|c|} \hline
    \multicolumn{2}{|c|}{Parameter} & BKS & FM & CHIK & $\epsilon_i$ \\ \hline\hline
    q$_{\rm{Si}}$ &[e]&  2.4 & 1.3743     & 1.9104   & 0.075   \\
    A$_{\rm{OO}}$ &[eV]&  1388.7730 & 666.3   & 659.5 &105   \\
    B$_{\rm{OO}}$ &[{\AA}$^{-1}$]&  2.76000 & 2.748     & 2.590 & 0.085     \\
    C$_{\rm{OO}}$ &[eV.{\AA}$^6$]&  175.0000 & 41.6    & 26.8 & 13.125    \\
    A$_{\rm{SiO}}$ &[eV]& 18003.7572 & 26486. & 27029. & 540 \\
    B$_{\rm{SiO}}$ &[{\AA}$^{-1}$]& 4.87318 & 5.1842     & 5.1586  & 0.03    \\
    C$_{\rm{SiO}}$ &[eV.{\AA}$^6$]& 133.5381 & 145.4   & 148.0  & 4.1 \\
    A$_{\rm{SiSi}}$ &[eV]& 0.0 & 3976.7  & 3150.4 & 300  \\
    B$_{\rm{SiSi}}$ &[{\AA}$^{-1}$]& 0.0 & 2.794     & 2.851 & 0.03     \\
    C$_{\rm{SiSi}}$ &[eV.{\AA}$^6$]& 0.0 & 882.5   & 626.7  & 66\\ \hline
  \end{tabular}
\caption{Parameters of the various potentials having the functional
form given by Eq.~(\protect\ref{BPOTENTIAL}). The last column gives
the values of $\epsilon_i$ used in Eq.~(\ref{eq6}) to calculate the
partial derivatives.}
\label{RGR_PARAM} 
\end{table}

Figure~\ref{potential_energy_shape}a gives the impression that the
three potentials are qualitatively quite similar, and that only their
absolute values are different. We point out, however, that it is not
possible to make the different potentials coincide by a simple shift
of the energy axis and we recall that 1\,eV corresponds to 11600\,K, thus
the differences are in fact remarkably large. This can also recognized
in Fig.~\ref{potential_energy_shape}b where we show the forces. In the
range between 2-3\,\AA\ these forces differ by up to a factor of two,
which shows that in fact the three force fields are very different.
Furthermore, it can also be noted that the values of the various parameters
for the three potentials are quite different: For example the charge of
a silicon atom differs by 20\%, and the prefactor $C_{\rm OO}$ of the van
der Waals term varies by a factor of 6. In view of these large variations,
we conclude that the problem of replacing the {\it ab initio}
potential by an effective two-body potential is somewhat ill-posed and
hence care has to be taken regarding the physical quantities (here forces
or structure) used for determining the optimal parameters. Hence, it is
important to check the quality of the obtained potentials by comparing
its predictions for quantities that have not been used in the fitting
procedure. This is done in the following section.

\section{Test of the potentials}
\label{sec3}

In this section, we will first compare the predictions of the two effective
potentials with the {\it ab initio} simulations results. As we will see,
the potential obtained from the force matching gives structures that are
rather different from the ones of the CPMD simulations and hence cannot
be considered to be accurate. Hence, in the second part of this section the
attention will be focused on the predictions of the CHIK potential, which
seems to be quite reliable, the obtained structures will be compared with
the ones predicted by the simulations of the CPMD and the BKS potentials.

\begin{figure}
\centering
\includegraphics*[width=0.43\textwidth]{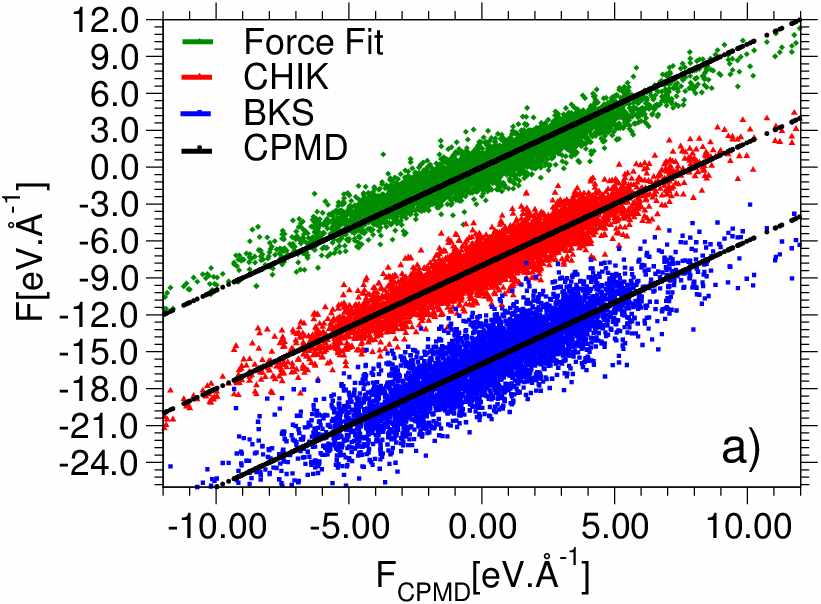}
\includegraphics*[width=0.43\textwidth]{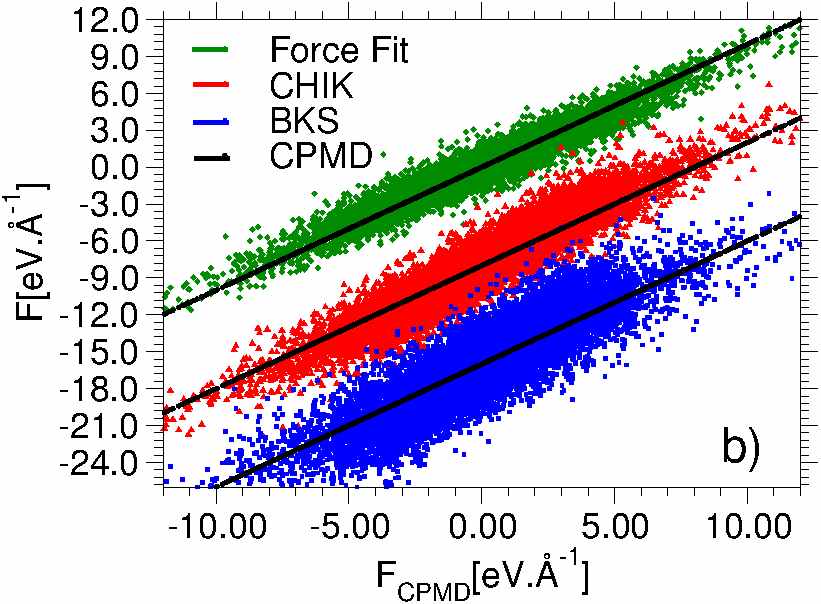}
\caption{
$x-$component of the force as obtained from the effective
force fields as a function of the $x-$component of the CPMD force. The
black straight lines are diagonals, \textit{i.e.}~correspond to a perfect
correlation. Panels a) and b) show respectively the Si and O forces. For
the sake of clarity the forces for the CHIK and BKS potentials have been
shifted downwards by 8 eV/\AA\; and 16 eV/\AA\;, respectively.}
\label{fig_force}
\end{figure}

To start, we compare how, for a given configuration, the forces as
predicted by the different potentials compare with the ones from the
CPMD simulations. In Fig.~\ref{fig_force}, we show a scatter plot of
the forces on a given particle as a function of the corresponding CPMD
force. (Here we show the force in the $x$-direction, but the ones in
$y$- and $z$-directions show of course qualitatively the same behavior). To
calculate this scatter plot we have used all 100 configurations
extracted from the CPMD run.
 If the effective potentials would indeed be able
to reproduce reliably the CPMD forces, all the points should lie on the
diagonal. We see that for the case of the FM potential the data points
are indeed quite close to the diagonal, which shows that this approach
is able to reproduce the CPMD forces with good accuracy. (Note that
although the FM potential has been optimized such that the forces are
well reproduced, it is {\it a priori} not evident that the functional
form given by Eq.~(\ref{BPOTENTIAL}) is indeed able to
give a good description of the forces).

Also included in Fig.~\ref{fig_force} is the data obtained from the
CHIK potential. We recognize that these forces deviate a bit more from
the diagonal than the ones from the FM potential, which is no surprise
since for the CHIK potential the information on the force was not taken
into account for the optimization of the parameters. Despite this larger
scattering, the correlation is still very satisfactory, giving thus
evidence that this potential is indeed reliable.

In this figure we have included also the forces derived from the BKS
potential and we see that these forces correlate significantly less
with the CPMD forces that the ones from the two other potentials. This
is somewhat surprising since in the past the BKS potential has been
found to be quite reliable in predicting the properties of amorphous
silica~\cite{Horbach_thesis}. Thus we can conclude that a good description
of the forces on a given atom is not a necessary condition for a potential
to give a good description of the material and later we will come back to
this point.  For the moment we thus just remark that it is satisfactory
to see that the CHIK potential is able to give a better description of
the forces than the BKS potential, even if this type of information
has not been used at all for the optimization of its parameters. To
quantify this trend we have calculated the $\chi^2_F$ as defined by
Eq.~(\ref{Chisq_Definition_Force_Fit}) but now we have used on the right
hand side the forces from the BKS and CHIK potentials. We have found
that the corresponding values of $\chi_F^2$ are respectively 1.22 and
0.54, thus showing that quantitatively the forces have indeed improved
significantly, even if they match the CMPD forces less good than the
one from the FM potential (for which the $\chi_F^2$ was 0.17).

From Fig.~\ref{fig_force}, it can also be observed that in the oxygen scatter plot for the
BKS as well as the CHIK potential the average slope is somewhat higher
than unity. This indicates that for these potentials the large forces
(in absolute value) are overestimated. This might be the reason why for
these effective potentials the vibrational properties at intermediate
and low frequencies are not very well reproduced~\cite{Benoit02}.

In order to test the reliability of the FM and CHIK potential we have
carried out a simulation of a $N=1152$ atoms system contained in a
cubic box of size $L=24.904$\,{\AA}, which corresponds to a density of
$2.2$\,g/cm$^3$. These simulations were done in the $NVT$ ensemble at
a temperature of 3600~K using an Andersen thermostat~\cite{Andersen}.
From the resulting configurations, we have then determined various
structural quantities which allowed us to make a direct comparison of
the structures obtained from the {\it ab initio} simulations with those
obtained from the ones with the effective force fields.

\begin{figure}
\centering
\includegraphics*[width=0.43\textwidth]{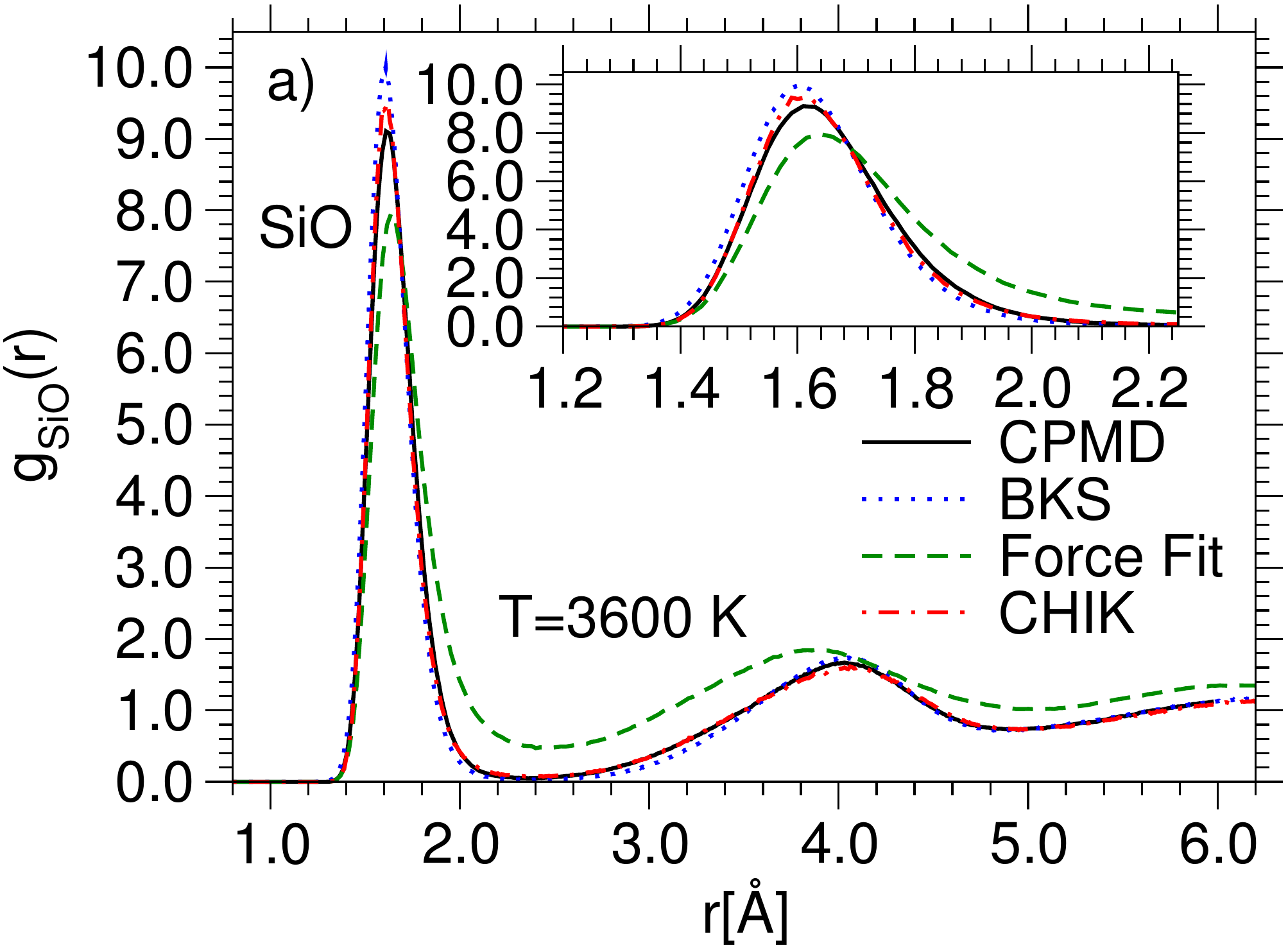}
\includegraphics*[width=0.43\textwidth]{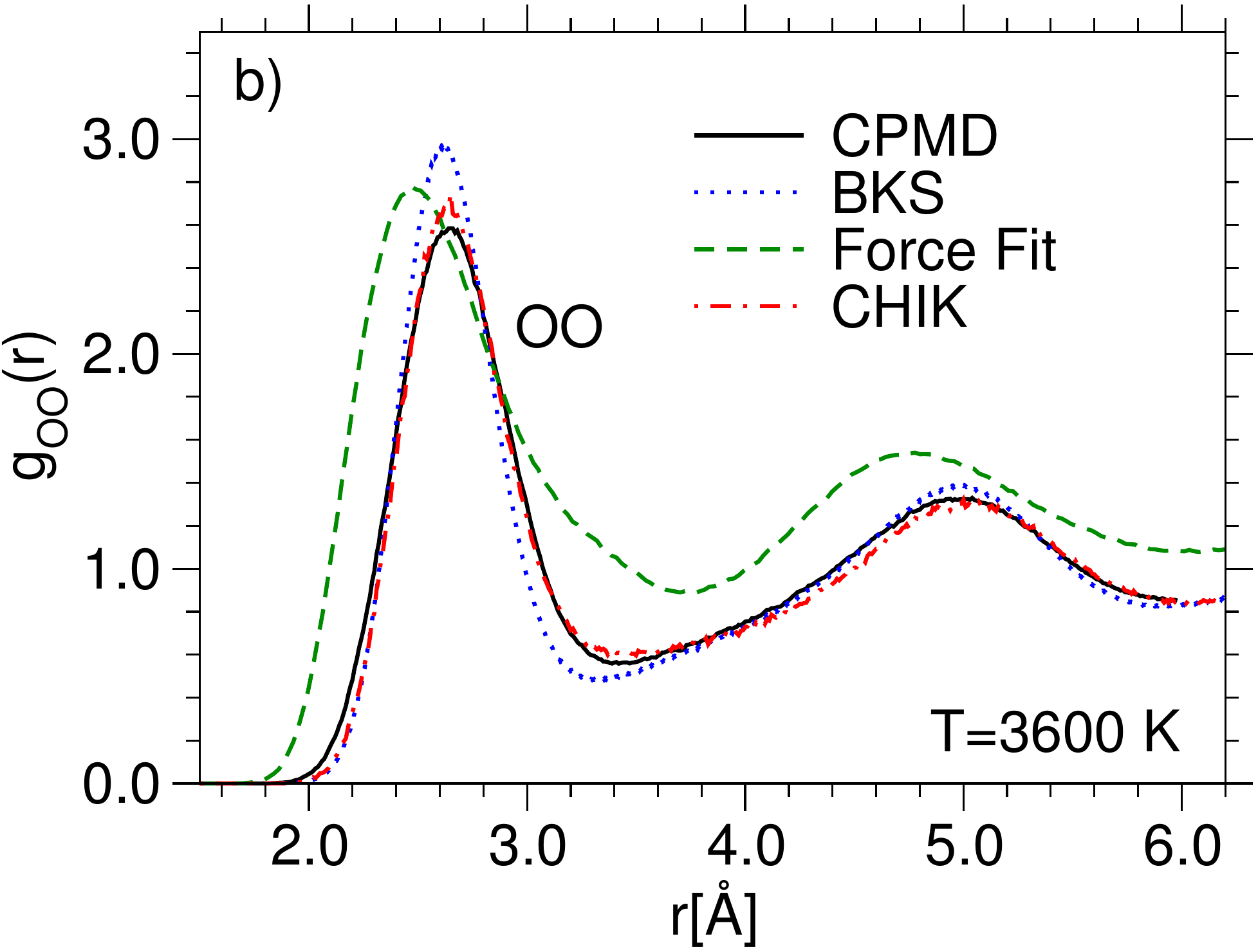}
\includegraphics*[width=0.43\textwidth]{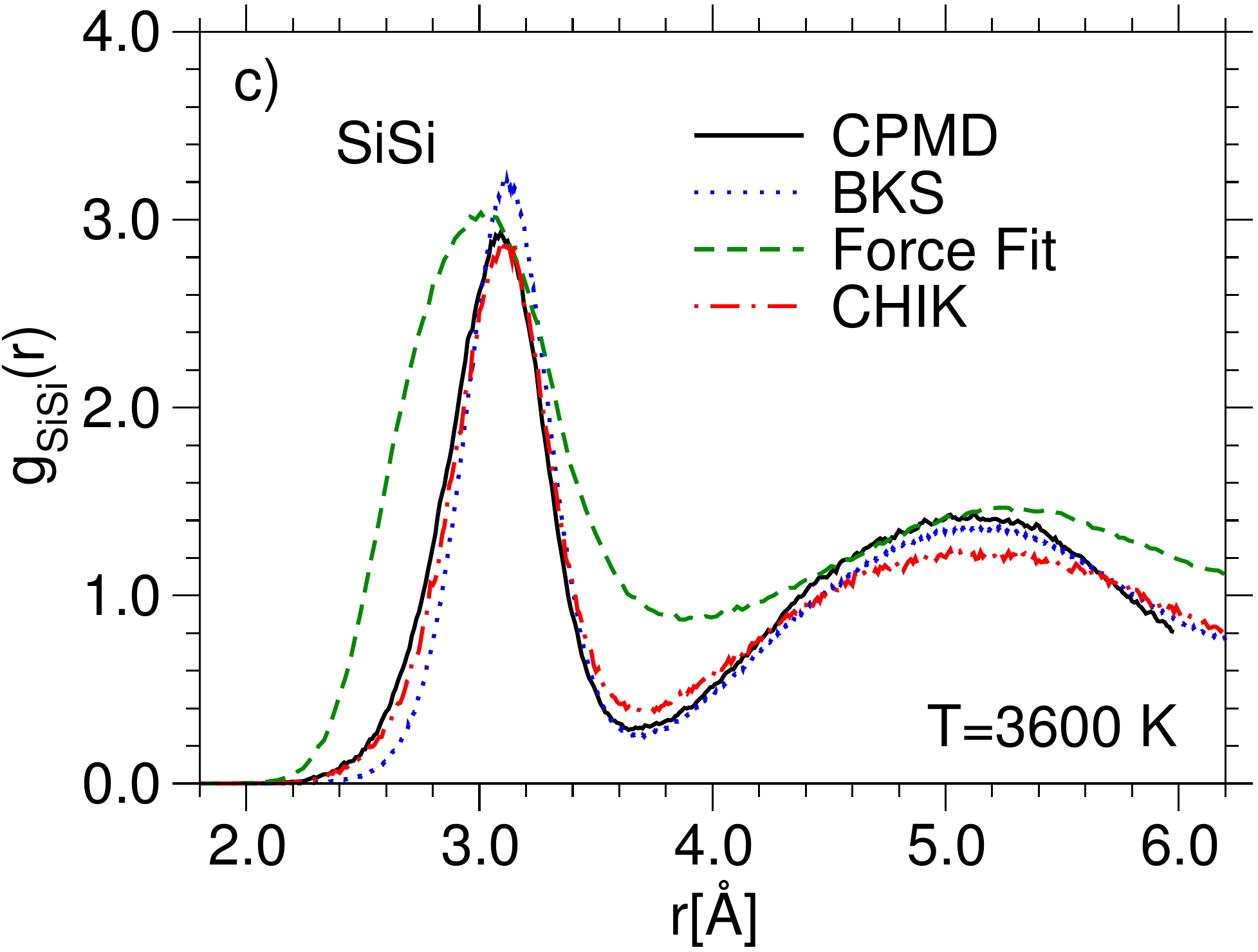}
\caption{
The different partial pair distribution functions at $T=3600$\,K. The
solid line corresponds to the \textit{ab initio} simulations. The dotted 
and dashed-dotted lines show the results for the BKS and CHIK potential,
respectively, and the dashed lines the one for the FM potential. The 
Inset in panel a) is a zoom onto the first peak.
}
\label{gr_CPMD_FF_BKS}.
\end{figure}

In Fig.~\ref{gr_CPMD_FF_BKS}, we show the pair correlation functions from
these simulations and compare them with the ones obtained from the CPMD
simulations. The Si-O correlation function, Fig.~\ref{gr_CPMD_FF_BKS}a,
shows that the BKS potential is able to give a good description of this
correlation in that this function is very similar to the one from the
CPMD. The main difference is seen in the first nearest neighbor peak
in that the BKS potential overestimates its height by about 10\%. For
this peak, the CHIK potential gives a better description, in that
the discrepancy is only about 5\%. We also see that this potential is
able to give an excellent description of the correlator in the vicinity
of the first minimum, whereas in that region the BKS
is slightly less
accurate.
For larger distances, the two effective potentials have a
comparable accuracy. The perhaps most astonishing result of that figure
is the curve for the FM potential in that we see that the correlator
matches the CPMD curve extremely poorly. Thus, we can conclude already at
this point that this potential is not able to reproduce the structure
in a reliable manner, and in the final section this result will be 
commented further.

In Fig.~\ref{gr_CPMD_FF_BKS}b we show the O-O correlation function. Again
we find that the curve from the BKS potential gives a good description
of the CPMD data, but overestimates the height of the first peak
and underestimates its width as well as the location of the first
minimum. More accurate is the correlator of the CHIK potential in that
the discrepancy regarding the height of the first peak is significantly
reduced and also the first minimum as well as the second nearest neighbor
peak are very similar to the one of the CPMD curve. The data from the
FM deviates again very strongly from the CPMD curve.

The Si-Si pair correlation function, associated with
the correlation that extends to the largest distance, is displayed
in Fig.~\ref{gr_CPMD_FF_BKS}c.
We see that the BKS potential gives
a rather good description of the second nearest neighbor peak, but
overestimates again the height of the first peak. The opposite trend
is found for the CHIK potential in that the first peak matches very
well the one from the CPMD simulations, but underestimates the height
of the second peak. Also here the curve from the FM potential strongly
deviates for all distances from the one of the CPMD simulation.

\begin{figure}
\centering
\includegraphics*[width=0.43\textwidth]{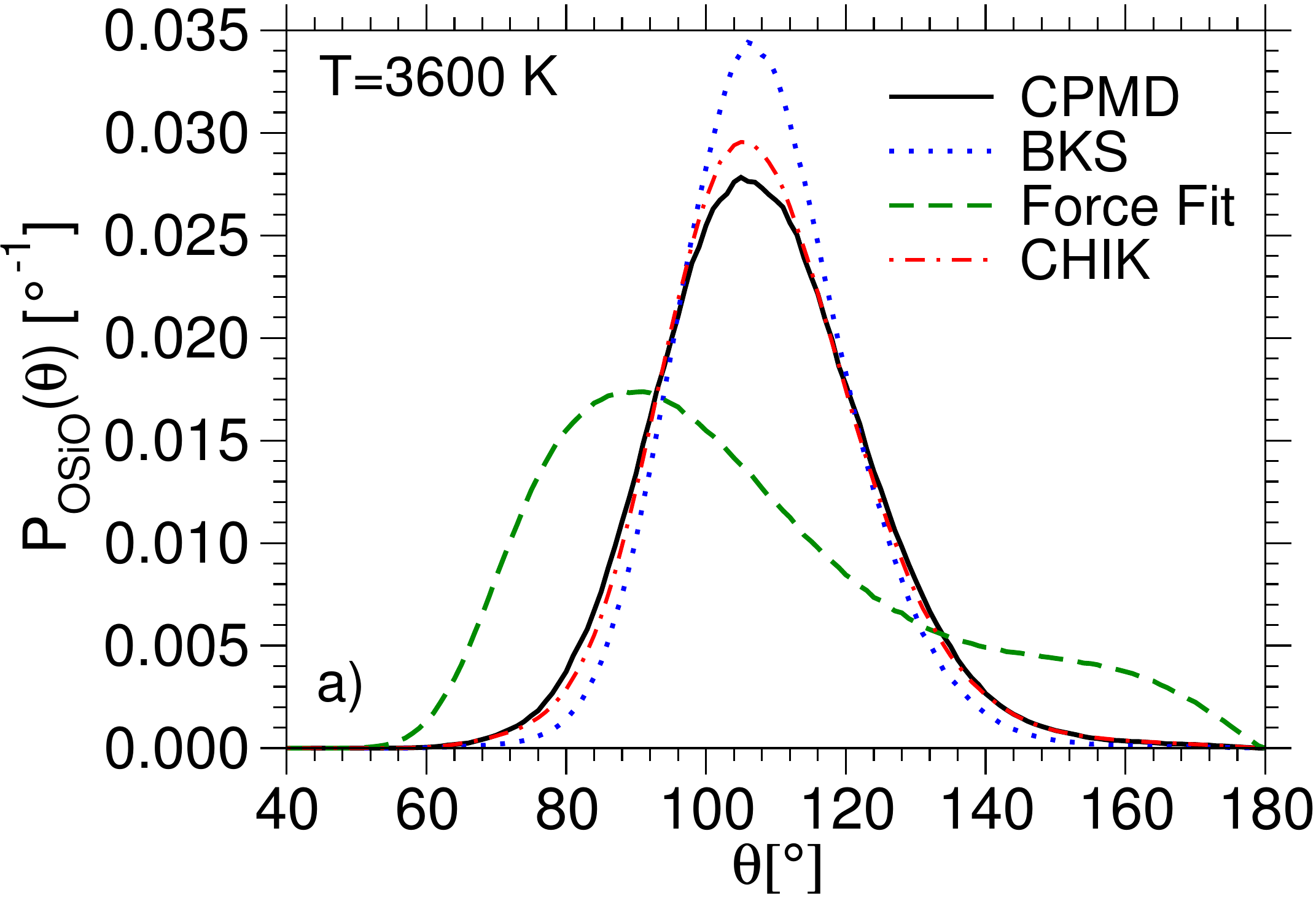}
\includegraphics*[width=0.43\textwidth]{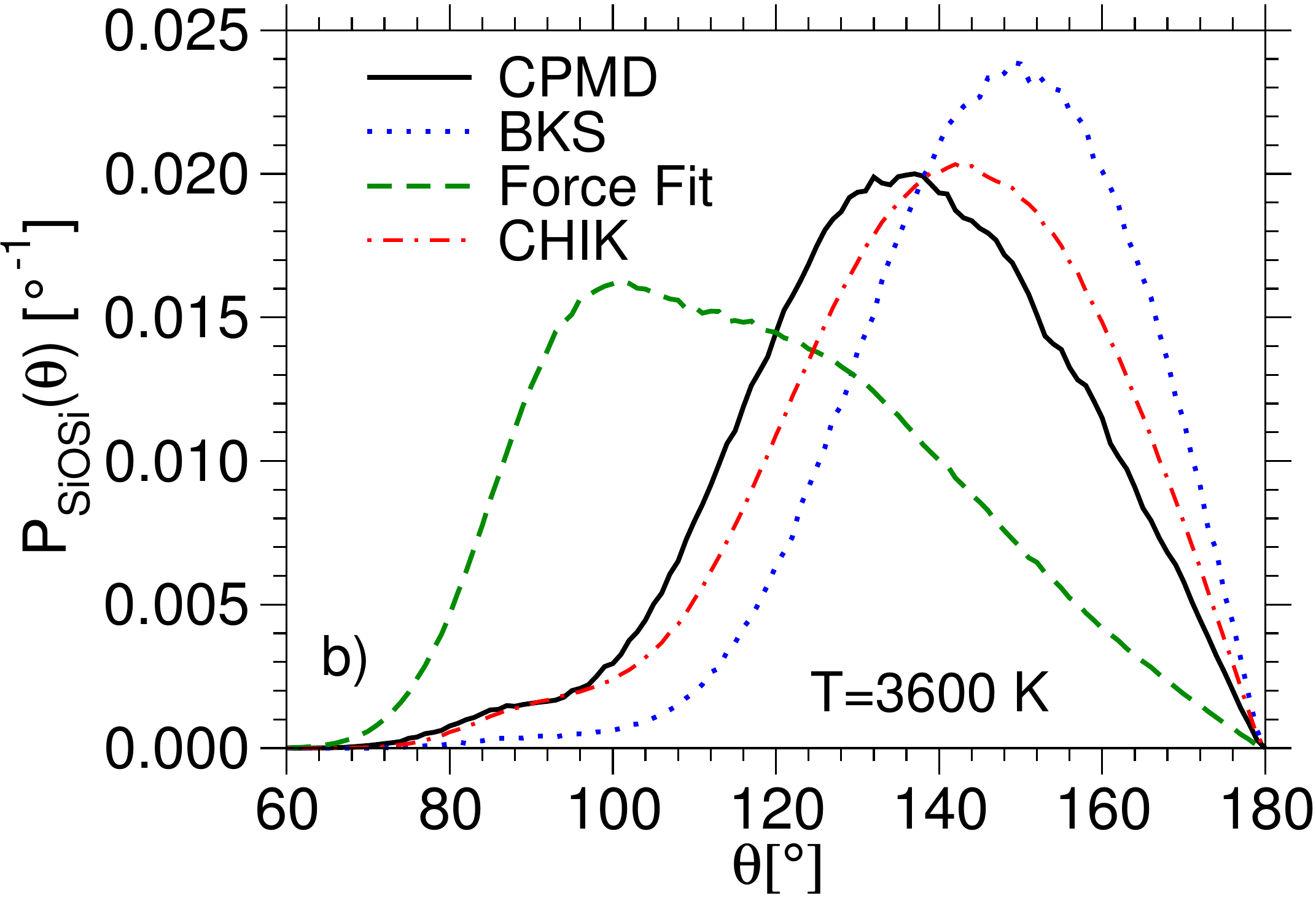}
\caption{Comparison of the angular distribution functions derived from \textit{ab initio}
simulation, BKS and CHIK simulations and simulations using the FM potential at $T=3600$\,K.
}
\label{angle_CPMD_BKS_FF}
\end{figure}

\begin{figure}
\centering
\includegraphics*[width=0.43\textwidth]{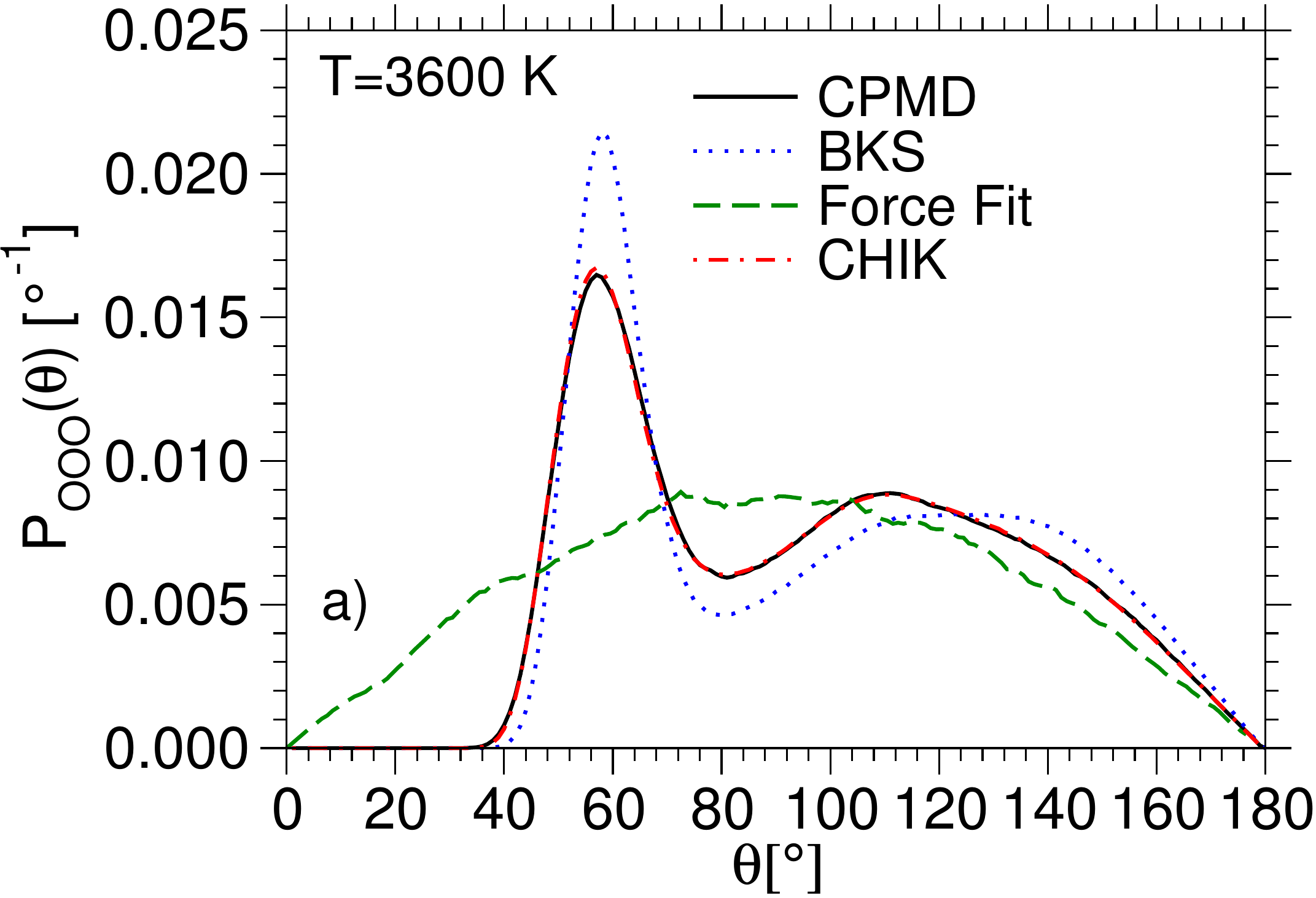}
\includegraphics*[width=0.43\textwidth]{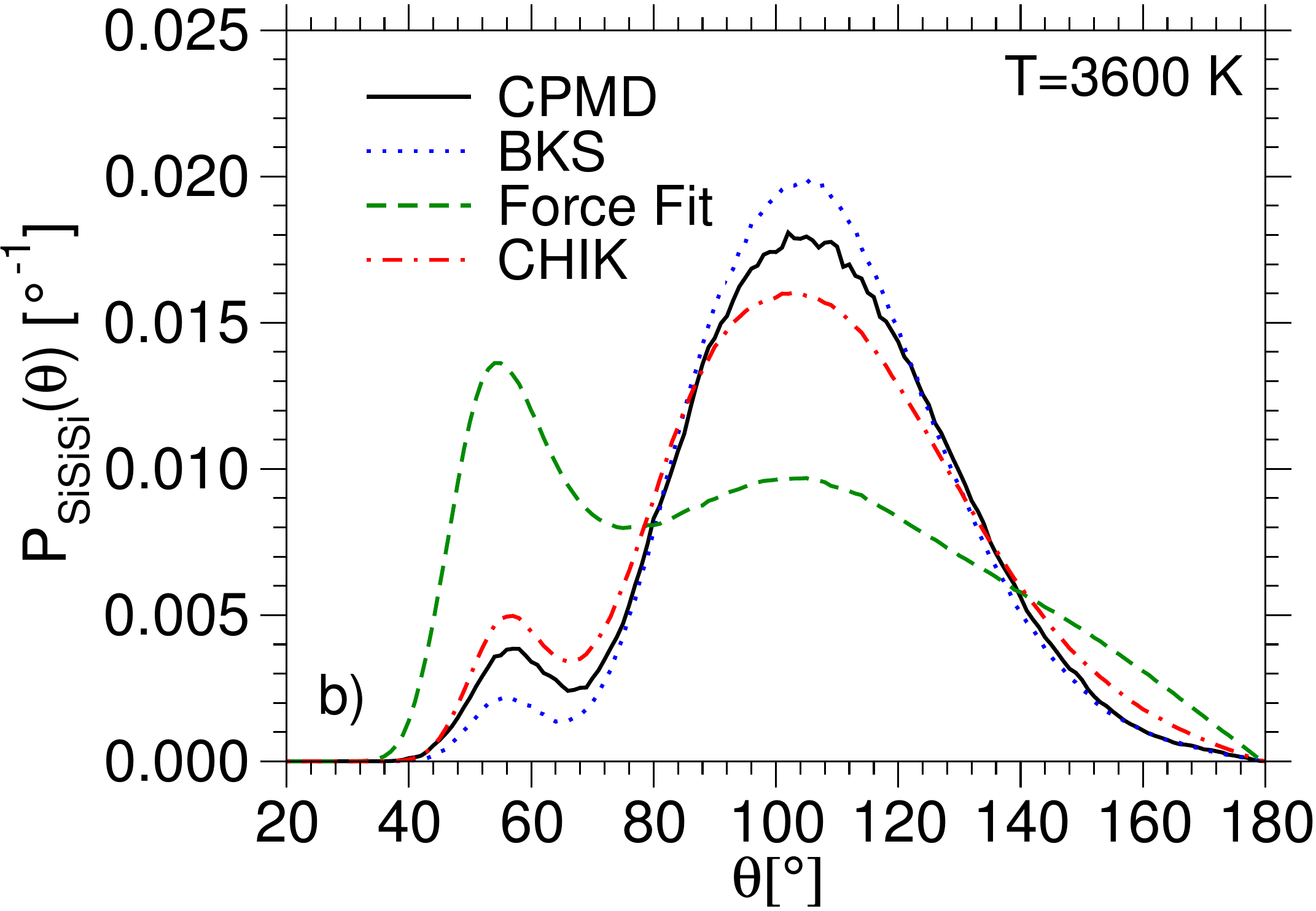}
  \caption{Angular distribution functions for the $\widehat{\mbox{OOO}}$ and
$\widehat{\mbox{SiSiSi}}$  angles at $T=3600$\,K computed using the 
CHIK and BKS potential and compared
to the CPMD results.\label{P_SiSiSi_OOO_T3600}.
}
\end{figure}

After having discussed the two body correlation functions, we now
present a comparison of the different potentials regarding a three
point correlation function, \textit{i.e.}~an observable that characterizes
the structure on a somewhat larger length scale. This is done by
considering the distribution of the angles between three neighboring
atoms, \textit{i.e.}~two atoms that are first nearest neighbor to the same
central atom. In Fig.~\ref{angle_CPMD_BKS_FF} and Fig.~\ref{P_SiSiSi_OOO_T3600}, we show the four
most relevant combinations of such triplets. The most important one
for the local structure is the angle $\widehat{\mathrm{OSiO}}$, see
Fig.~\ref{angle_CPMD_BKS_FF}a, since it determines largely the
shape of the tetrahedra. We note that the BKS potential as well as
the CHIK potential reproduce well the peak around $106^\circ$, \textit{i.e.}~the
typical angle in a tetrahedron, predicted by the CPMD simulations. The
BKS potential overestimates the height of the peak by about 10\%
whereas for the CHIK potential this discrepancy is reduced to 6\%. The
distribution from the FM potential is very different from the one of the
CPMD simulations and we conclude that this potential does not give rise
to a realistic structure.

The angle $\widehat{\mathrm{SiOSi}}$ characterizes the relative
arrangement of two neighboring tetrahedra. The corresponding distribution
function is shown in Fig.~\ref{angle_CPMD_BKS_FF}b and we see that
for this quantity the CHIK potential is significantly more reliable
than the BKS potential in that the position and height of the main peak
of the former are much closer to the one of the CPMD distribution than
the ones of the latter. Also for this case the distribution for the FM
potential is very different from the one of the CPMD.

The angle $\widehat{\mathrm{OOO}}$ gives information on the
shape of a tetrahedron, as well as the relative orientation
of two neighboring tetrahedra, and its distribution is shown in
Fig.~\ref{P_SiSiSi_OOO_T3600}a.  The first peak at $\theta=58^\circ$
corresponds to the angles made by three oxygen atoms belonging to the
same tetrahedron. The height of this peak is significantly overestimated
by the BKS potential indicating that this potential tends to generate
structures that are too regular.  As a consequence the second peak,
located at $\theta\simeq120-140^\circ$ and related to the oxygens on two
neighboring tetrahedra, is lower and also flatter than the one predicted
by the CPMD simulations. What is surprising in the figure is that the
CHIK potential gives an almost perfect description of this distribution
function. In contrast to this, the distribution from the FM potential
is not good at all.

Finally, we show in Fig.~\ref{P_SiSiSi_OOO_T3600}b the distribution of
the angle $\widehat{\mathrm{SiSiSi}}$, \textit{i.e.}~the angle between
three neighboring tetrahedra. The graph shows that the BKS as well
as the CHIK potential give a good although not perfect description of
this distribution function. The position of the small peak at around
$\theta \simeq 56^\circ$, which is due to three-membered rings see 
Ref.~\cite{Rino}, is reproduced well,
but its height is over/underestimated by the CHIK and BKS potentials
respectively. The main peak at around $\theta \simeq 104^\circ$ is
due to larger rings and its height is again not reproduced very well,
although its position is. Also in this case the prediction of the FM
potential is not reliable.

So far we have checked to what extent the two effective potentials
are able to reproduce the {\it ab initio} data at $T=3600$~K, \textit{i.e.}~at
the temperature for which the various parameters characterizing these
potentials have been optimized. However, it is of course also important to
test whether these potentials are also capable to make reliable predictions
at different temperatures, notably in the glass state. In the following
we will thus present such tests. Since we have seen that the
FM potential does a rather poor job in reproducing the CPMD data, for
reasons that will be discussed in the final section, we will carry out
these tests only for the CHIK potential and compare them with the results
from the BKS potential.

\begin{figure}
\centering
\includegraphics*[width=0.43\textwidth]{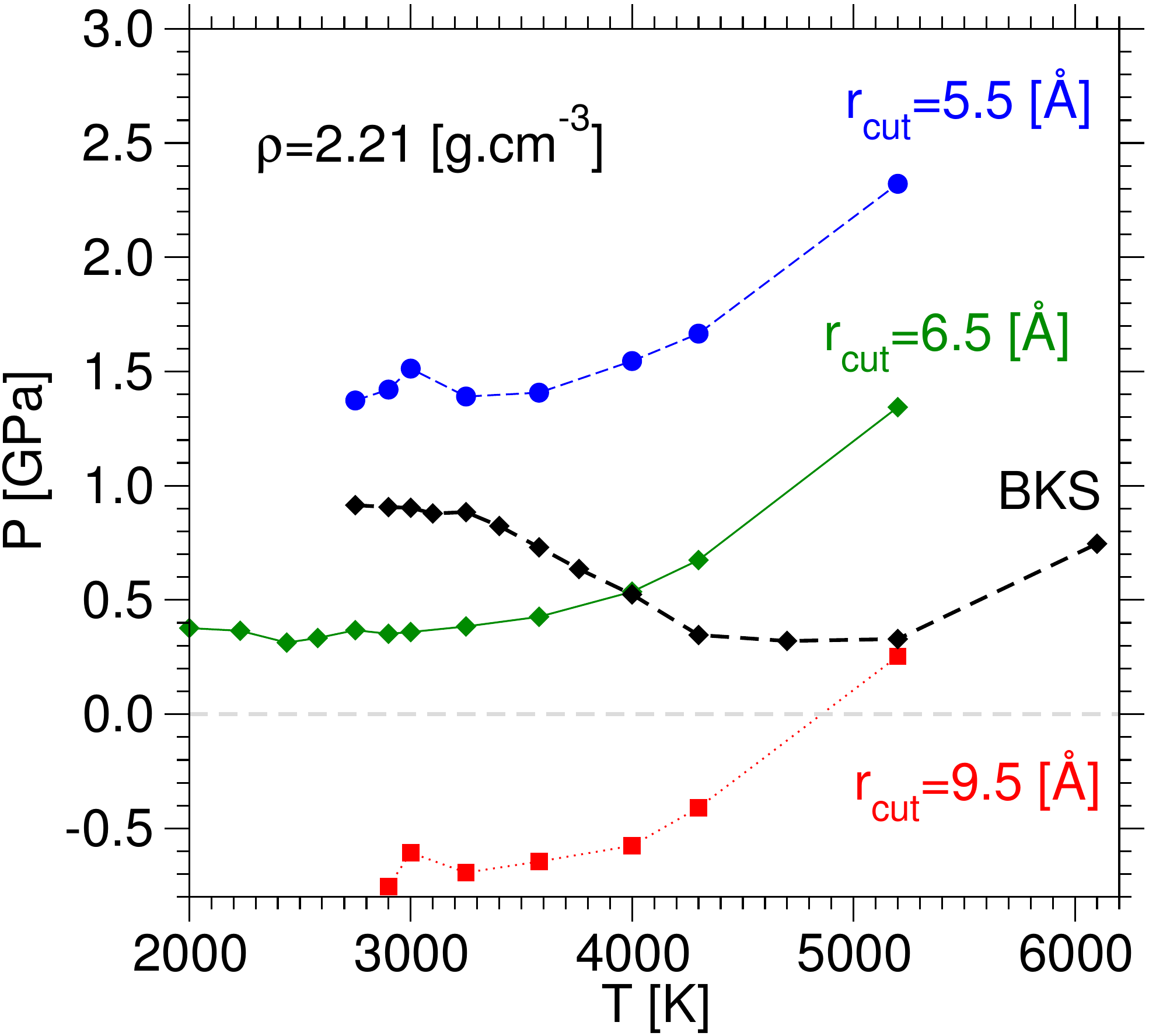}
\caption{
Temperature dependence of the pressure as predicted by the BKS potential
and for the CHIK potential with different cut-off radii for the short
range part of the potential.}
\label{P_fig}
\end{figure}

To this end, we start with the equation of state by monitoring how the
pressure $P$ varies as a function of temperature if the density is
kept fixed.  We recall that silica has an anomaly in the density in that
upon heating from low temperatures at constant pressure, its density
first {\it increases} and only at intermediate and high temperatures
it starts to decrease (or, if one considers heating of the system at constant
volume, the pressure first decreases, before it starts to increase). In
experiments, this anomaly, which is related to the open network structure
of the system, is observed at around 1800\,K~\cite{bruckner}. In
Fig.~\ref{P_fig} we show the temperature dependence of $P$ for the
case of the BKS potential. In agreement with previous simulations
at constant pressure~\cite{vollmayr96}, one finds an anomaly at
around 4800\,K, \textit{i.e.}~the pressure dependence is qualitatively correct,
but quantitatively the agreement is not that good. Also included in
the figure is the result for the CHIK potential. We see that also this
potential predicts an anomaly, although it is not very pronounced. On the
other hand, the temperature at which this anomaly occurs is now shifted
down to around 2200\,K, \textit{i.e.}~relatively close to the temperature at which
experiments show the maximum in the density.

Since Vollmayr \textit{et al.}~have found that the pressure depends
quite strongly on the cut-off distance for the Buckingham terms of
the interatomic potential~\cite{vollmayr96}, it is of interest to
check how this distance influences the equation of state for the CHIK
potential. The result is included in Fig.~\ref{P_fig} as well in that we
show data for different values of this cut-off distance.  We observe
that changing this distance leads to a vertical shift of $P(T)$ that is
basically independent of $T$. In particular we find that a decrease of
$r_{\rm cut}$ by 1\,\AA~leads to an increase of the pressure of about
1\,GPa, whereas an increase of 2\,\AA~makes that the pressure decreases by
about 1\,GPa. Thus in summary we can conclude that the choice of $r_{\rm
cut}=6.5$\,\AA\; is a very reasonable one.

Next, we study the structure of glasses produced with the various
potentials. For this we have equilibrated a system of $1152$ atoms with the
CHIK potential at $T=2440$\,K and subsequently quenched it to $T=300$\,K
using a very high quench rate. To improve the statistics of the results
we have repeated this procedure for 16 independent initial configurations.

We will compare the results with the data from three {\it ab initio} samples,
discussed in Ref.~\cite{Benoit02}, that have been obtained by quenching
samples of the BKS potential to $300$\,K and subsequently annealing them
by means of a CPMD run of $0.7$\,ps. Furthermore, we will also present
data for the BKS potential. These glass structures have already been
studied in Ref.~\cite{vollmayr96} and have been obtained by cooling at
constant pressure ($P=0$\,GPa) ten systems of 1002 atoms with a quench
rate of $4.44\times10^{12}$\,K/s to zero temperature.

\begin{figure}[th]
\centering
\includegraphics*[width=0.43\textwidth]{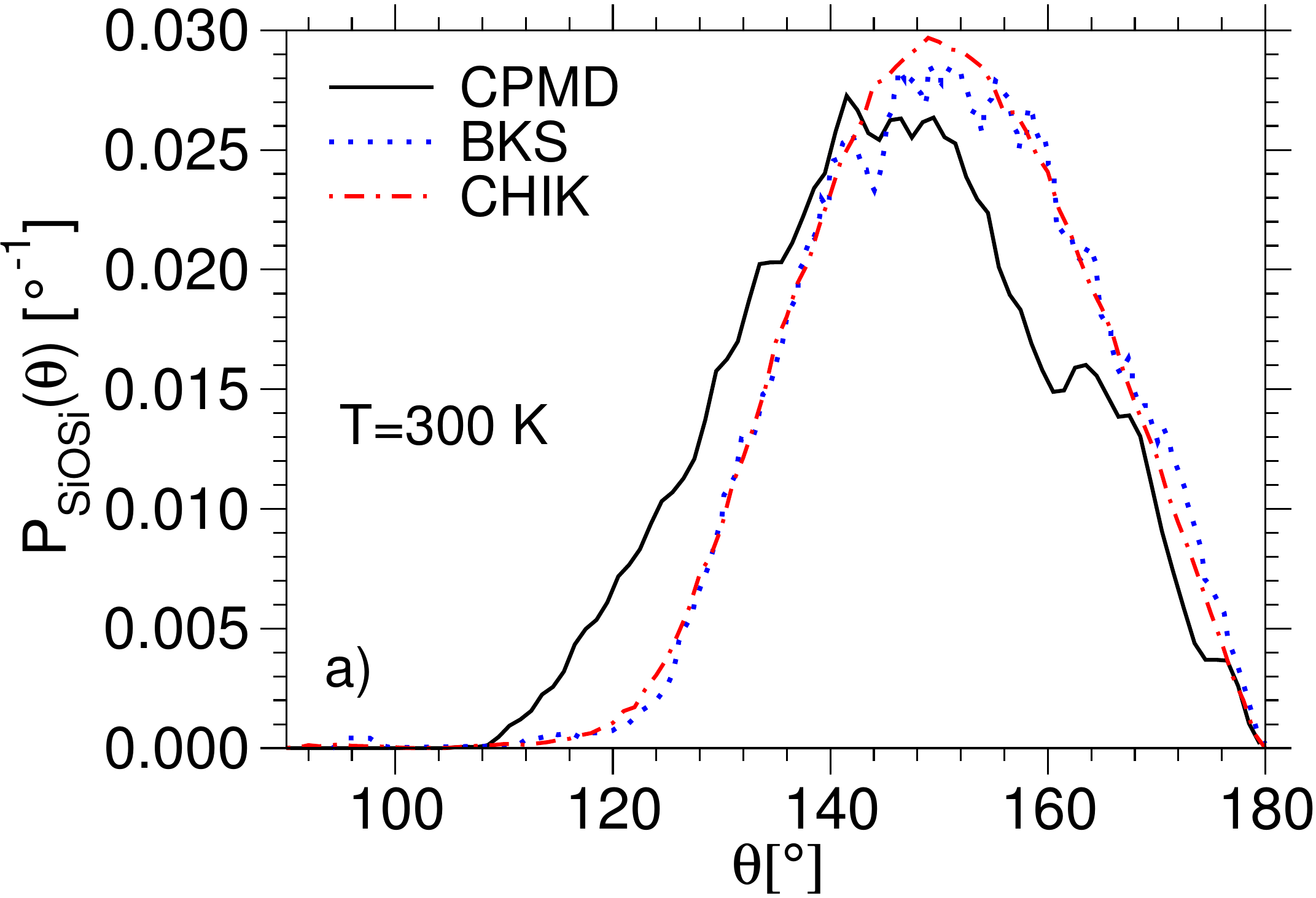}
\includegraphics*[width=0.43\textwidth]{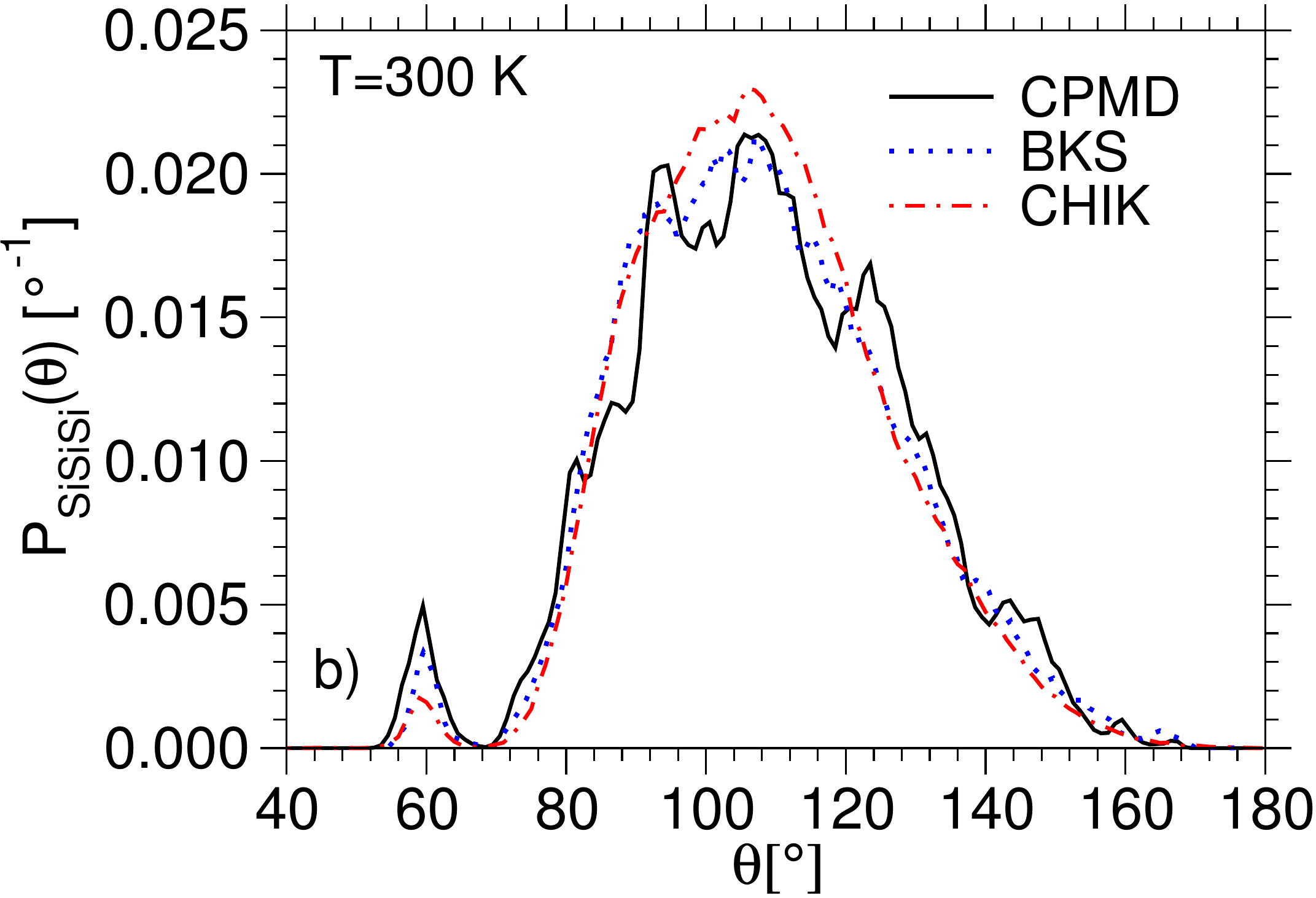}
\includegraphics*[width=0.43\textwidth]{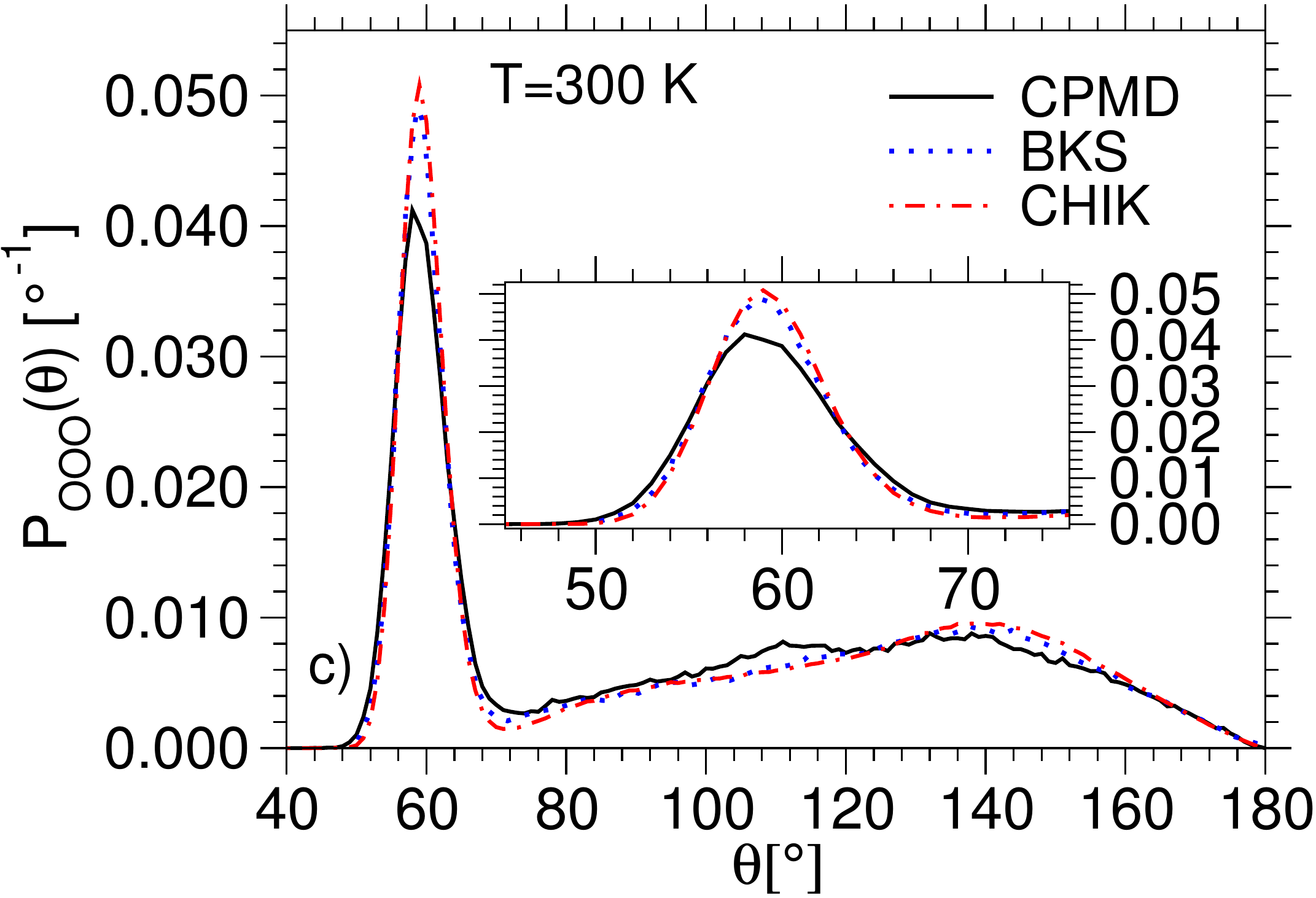}
\caption{Angular distribution functions for the $\widehat{\mbox{SiOSi}}$,
$\widehat{\mbox{SiSiSi}}$, and $\widehat{\mbox{OOO}}$ angles at $T=300$\,K
as obtained from the CHIK and BKS potentials and compared to the CPMD
results. }
\label{P_SiSiSi_OOO_T300}
\end{figure}

Before we start with the comparison of the glass properties, we
recall that these features do depend on the quench rate with which
the glasses have been produced~\cite{vollmayr96}. Therefore, one has to
be careful if one compares the properties of glass samples that have been
produced using very different cooling rates. However, these cooling rate
effects are relatively small, depending on the observable considered
on the order of 1-10\%~\cite{vollmayr96}, and hence are significantly
smaller than the variation one gets by using different potentials (see
below). Thus, in the following we will not discuss these cooling-rate effects.

In Fig.~\ref{P_SiSiSi_OOO_T300}, we show the angular distribution
functions as obtained from the three different potentials. The
distribution of the $\widehat{\mbox{SiOSi}}$ angle is shown in panel~(a)
and we see that the one predicted by the BKS potential basically
coincides with the one of the CHIK potential in that they both have
a maximum at around 150$^\circ$. A comparison with the corresponding
distributions in Fig.~\ref{angle_CPMD_BKS_FF} shows that the
different maxima have shifted by $\Delta \theta^{\mathrm{CP}}\simeq
9^\circ$, $\Delta \theta^{\mathrm{BKS}}\simeq 4^\circ$, and $\Delta
\theta^{\mathrm{CHIK}}\simeq 7^\circ$, indicating that the CHIK
potential has a stronger $T-$dependence of the average angle than
the BKS potential. Note that the experimental value for this angle,
obtained from NMR data~\cite{Mauri}, is around 151$^\circ$, thus it is
in quite good agreement with the angle predicted by the CHIK and BKS
potential. Surprisingly, this value is, however,  larger
than the one predicted by the CPMD data, $\theta^{\rm CP}_{\rm max} \simeq
145^\circ$. Thus, at this point one might start to wonder whether
the CPMD simulation used to relax the glasses generated by the BKS
potential~\cite{Benoit02,Benoit_Ispas_1} are indeed able to anneal the structure
in a significant manner. More studies on this point would certainly
be very useful. Thus, if we assume that the distribution of this angle
as predicted from the CPMD is not quite reliable and instead take the
experimental value for the mean angle as a reference, we can conclude
that the CHIK potential is indeed able to give a surprisingly good
description of this angle for the glass at room temperature.

For the distribution of the angle $\widehat{\mbox{SiSiSi}}$, shown
in Fig.~\ref{P_SiSiSi_OOO_T300}b, we see that the BKS as well as the
CHIK potential give results that are in good agreement with the CPMD
data. However, due to the lack of statistics the latter is rather noisy
and therefore it is at this point not possible to conclude much more
from this agreement. 

Regarding the $\widehat{\mbox{OOO}}$ distribution, shown in
Fig.~\ref{P_SiSiSi_OOO_T300}c, we can conclude that also here the
agreement between the two effective potentials and the CPMD data is good,
although both overestimate somewhat the height of the peak at around
60$^\circ$. Since we have seen in Fig.~\ref{P_SiSiSi_OOO_T3600}b that at
3600\,K the CHIK potential shows an excellent agreement with the
CPMD data, we can thus conclude that this potential is able to give a good
description of this angular distribution at several temperatures.

In experiments on atomic systems it is not possible to access in a
direct way the partial pair distribution functions discussed above
(Fig.~\ref{gr_CPMD_FF_BKS}). Instead one can sometimes measure
their space-Fourier transforms, \textit{i.e.}~the partial structure factors
$S_{\alpha\beta}({\bf q})$, where ${\bf q}$ is the wave-vector. However,
in most cases these partial observables are also not accessible and only the
neutron scattering function $S_n({\bf q})$, which is a weighted average
over these partials, can be measured. This function is given by \cite{Lovesey}
\begin{equation}
  S_n({\bf q})= \frac{1}{N_{\rm{Si}} b_{\rm{Si}}^2 +N_{\rm{O}} b_{\rm{O}}^2}
  \sum_{j=1}^{N} \sum_{k=1}^{N} b_j b_k \left \langle \exp \left[ i {\bf q}
      \cdot ({\bf r}_j-{\bf r}_k) \right]  \right \rangle \label{NSQ} .
\end{equation}
Here the coefficients $b_{\alpha}$ with $\alpha \in \{\rm{Si,O}\}$ are the
neutron scattering cross sections, $N_{\alpha}$ is the number of atoms
of type $\alpha$ and $\langle \cdot \rangle$ is the thermal average.
To calculate $S_n({\bf q})$ from the partial structure factors we have
used the experimental values $0.4149 \times 10^{-12}$\,cm and $0.5803
\times 10^{-12}$\,cm for $b_{\rm{Si}}$ and $b_{\rm{O}}$, respectively
\cite{nist}.

\begin{figure}
\centering
\includegraphics*[width=0.43\textwidth]{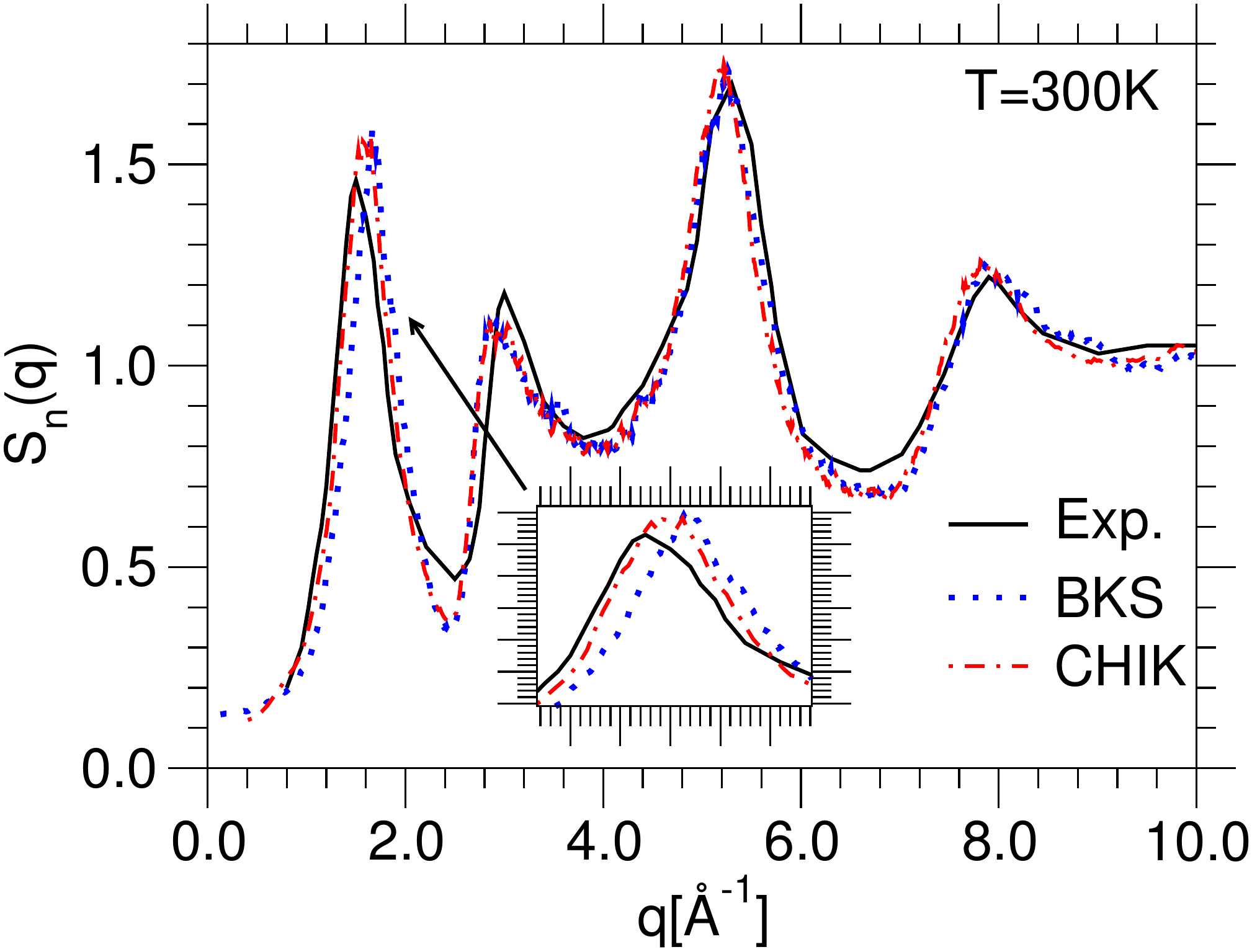}
\caption{Comparison of the neutron scattering function from our simulation
(red dashed line) with the experimental data of Susman \textit{et al.}
\cite{Susman} (solid black line) and with simulation data with the BKS potential~\cite{horbach99}
(blue dotted line). The Inset shows a zoom on the first peak.
}
\label{sq_fig}
\end{figure}

The structure factor thus obtained from the different potentials
is shown in Fig.~\ref{sq_fig}. In this figure, the data from the BKS
potential has been taken from Ref.~\cite{horbach99} and the experimental data
from Ref.~\cite{Susman}.

From this figure we see that the simulations with the effective
potential are in quite good agreement with the experimental data. For the
CHIK potential the position of the first peak, related to the nearest
neighbor distance between two tetrahedra, reproduces the experimental
value better than the BKS potential (see Inset), a trend that is very reasonable if
one recalls the results on the radial distribution functions discussed
above. Finally, we also note that the first peak, as well as the first
minima are slightly over- and underestimated, respectively, indicating
that the simulated structures are a bit too regular.

\subsection{Vibrational density of states}

So far we have mainly studied to what extent the effective potentials
are able to reproduce the structural properties of the liquid and
glass. However, many quantities that are relevant for practical
applications do not depend on the geometry of the structure, but
instead on the vibrational properties of the glass. Important examples
for such quantities are the specific heat, the optical transmission
coefficient, etc. It is therefore of interest to check whether the
effective potentials are also able to reproduce the vibrational spectra,
also called vibrational density of states (vDOS), of the system. We recall
that this spectrum is given by the eigenvectors of the dynamical matrix,
which is in turn directly related to the Hessian matrix, $\partial^2
V/\partial {\bf r}_i \partial {\bf r}_j$. Thus, the vDOS can be considered
as a measure on how accurate the potential is able to describe the
local curvature of the potential in the vicinity of a local minimum,
\textit{i.e.}~a glass state.

Previous studies on silica glasses have shown that {\it ab initio}
simulations are able to reproduce quite reliably the vDOS as obtained from inelastic neutron scattering experiments
\cite{Benoit02,Sarnthein,giacomazzi,haworth}. This is also the case for the silica
considered here and hence we can use the vDOS from the CPMD calculation
as reference data.

\begin{figure}
\centering
\includegraphics*[width=0.43\textwidth]{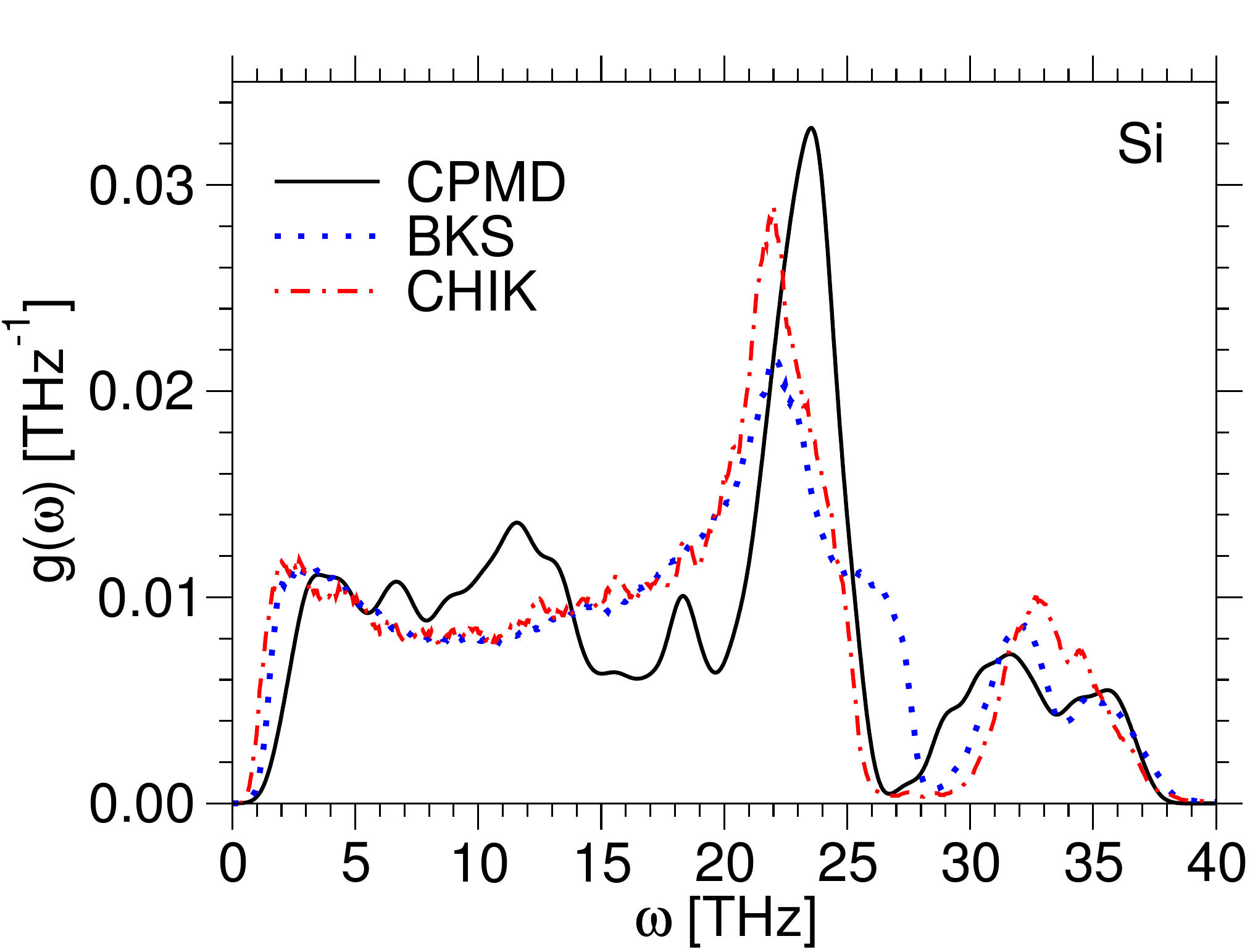}
\includegraphics*[width=0.43\textwidth]{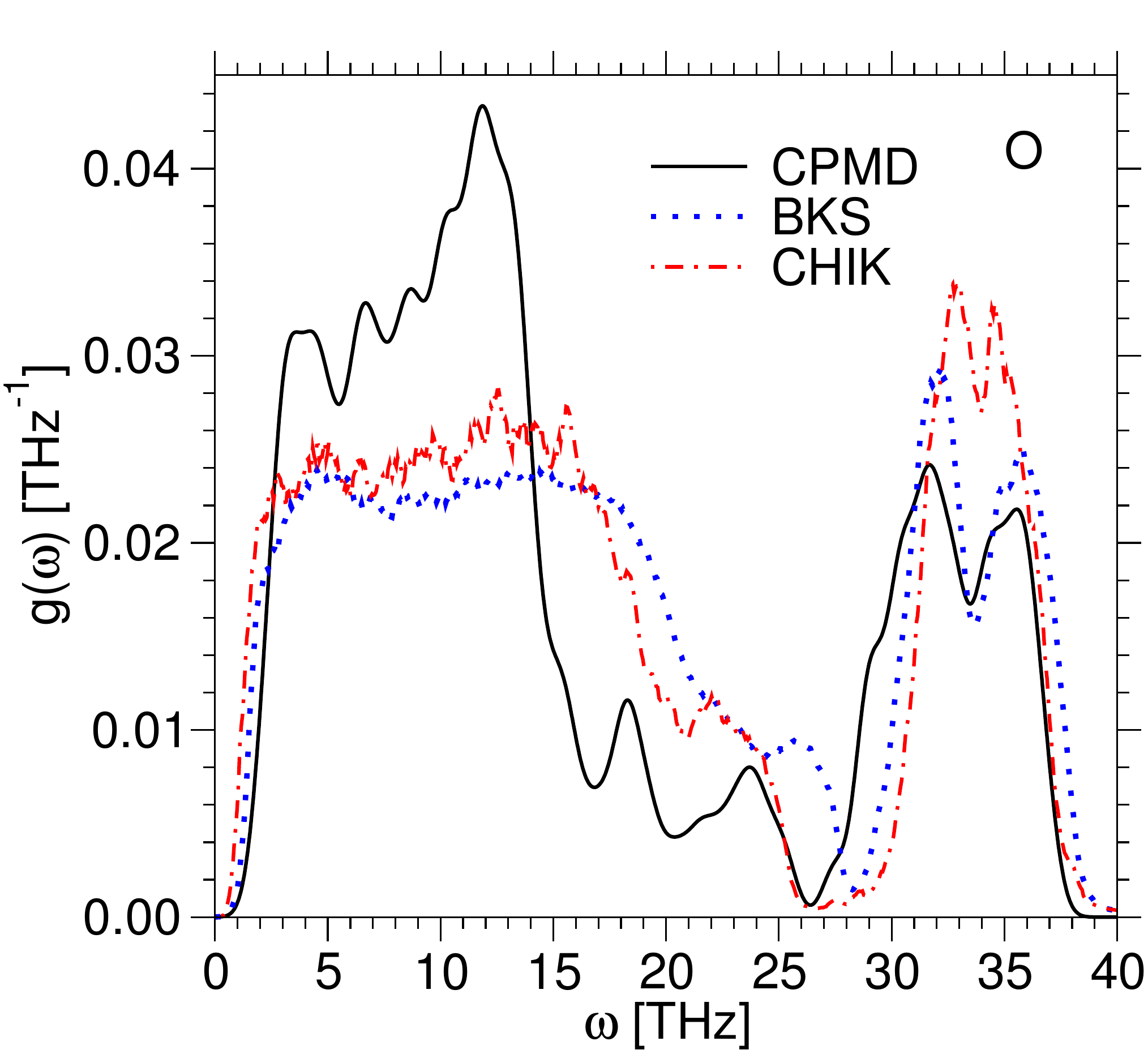}
\caption{
Partial vibrational density of states.  The curves corresponding to
the CHIK and BKS potentials have been determined
using the velocity autocorrelation function at $T=300$\,K on a sample
with $N=1152$ and $N=8016$ atoms~\cite{horbach99}, respectively. The \textit{ab initio}
data have been obtained at $0$\,K by calculating the eigenvalues of the
dynamical matrix on  three $N=78$ atoms samples and by convoluting the resulting spectrum with a
Gaussian broadening of width $2 \sigma = 1.05$\,THz.\label{DOS_fig}}
\end{figure}

In practice, we have determined the vDOS $g(\nu)$, where $\nu$ is the
frequency, by using the glass configurations generated with the CHIK
potential at 300\,K (and from which we have obtained the structure of
the glass discussed above) and started a standard $NVE$ simulation. By
recording the velocity of all the particles, we calculated the Fourier
transform of the mass-weighted velocity auto-correlation function. Within
the harmonic approximation, which can be expected to hold at temperatures
well below the glass transition temperature, this Fourier transform is
then directly related to the vDOS via \cite{Dove}
\begin{equation}
g(\nu) = \frac{1}{N k_B T} \sum_j m_j \int_{-\infty}^{\infty} dt \langle {\bf
  v}_j(t) \cdot {\bf  v}_j(0) \rangle \exp (i 2 \pi \nu t).
\end{equation} 
Note that by restricting in this relation the sum to only one species
of particles, one can obtain the so-called partial vDOS which
gives information on the vibrational behavior of a given species.
The so obtained vDOS can be compared to the \textit{ab initio} data
published in Ref.~\cite{Benoit02} and to the classical
data obtained from the BKS potential \cite{vollmayr96,Horbach_thesis},
see Fig. \ref{DOS_fig}.

For the high frequency domain, $\nu>25$\,THz, 
Sarnthein \textit{et al.}~have shown that the observed doublet is due to local vibrational modes of
the SiO$_4$ tetrahedra, with the peak at $\nu=31.5$\,THz being related
to an out-of-phase motion of oxygen atoms in which two oxygen
atoms move towards the central Si atom while the other two move away,
and the peak at $\nu=35$\,THz to an in-phase motion of all oxygen
atoms toward the Si atom \cite{Sarnthein_Pasquarello}.
As can be seen in Fig.~\ref{DOS_fig}, the BKS predicts the location and
intensity of these two peaks more accurately than the CHIK potential. This
is not that surprising since in the parameterization of the BKS potential
one has made use of the information of the vibrational modes located
within one tetrahedron~\cite{bks}. In contrast to this, the CHIK
potential does not have this kind of input and it is therefore a
positive surprise that it is nevertheless able to reproduce these two
peaks with a quite good quality.

The frequency range below $25$\,THz shows a broad band that includes
several peaks. We see that the CHIK potential gives a very good
description of the high frequency edge of this band, and in particular it
is in this $\nu-$range significantly more accurate than the BKS potential.
The peak around $24$\,THz, very pronounced in the CPMD data, is due to
the bending of the Si-O-Si angle in the  Si-O-Si plane~\cite{Pasquarello98}.
Although the CHIK potential underestimates somewhat the frequency of these
excited states, it gets its intensity about right, and also in this case it
fares better than the BKS potential.

For lower frequencies the vDOS of both classical potentials is relatively
featureless and they are in fact very similar. This is in contrast to
the CPMD data which show various peaks: The one at around 18\,THz is
related to modes associated with three-membered rings, and the one
at around 12\,THz to four-membered rings~\cite{Pasquarello_Car}. In
agreement with previous studies, the classical potentials are not able
to reproduce these spectral features and at present no obvious reason
is known why this is so~\cite{huang15}.

\section{Summary and Conclusions \label{CAO}}

The goal of the presented work was to develop a method in which one
uses the microscopic information provided by {\it ab initio} simulations to
develop effective classical potentials. In addition, we wanted to compare
our approach with the force matching method proposed by Ercolessi
and Adams~\cite{ercolessi94} since for certain systems this approach has given
satisfactory results.

In the present approach, we started from the idea that if the effective
potential is able to reproduce reliably the local arrangement of the
atoms, it should give a good description of many other properties of the
systems as well. Therefore, we have used the {\it ab inito} simulations
at $T=3600$\,K to calculate the partial radial distribution functions and
then defined a cost function, Eq.~(\ref{chi_def}), that characterizes
how accurate an effective potential can reproduce these distribution
functions. In a first step we have tested what choice should be done to
define ``local'' by varying the exponent $n$ in Eq.~(\ref{chi_def})
and found that $n=1$ is a good compromise since it gives a substantial weight
to the structure at the first and second nearest neighbor distance.

Using this cost function we have optimized the parameters of the
functional form given by Eq.~(\ref{BPOTENTIAL}) and obtained a new
interaction potential (referred to as CHIK) for amorphous SiO$_2$. A
comparison of the predictions of this new potential with the
ones of the well studied BKS potential shows that for most structural
properties the CHIK potential gives a better description of the CPMD
data than the BKS potential, and that sometimes the improvement is quite
significant. This superiority is found also in the equation of state in
that the CHIK potential predicts a density anomaly at a temperature that
is closer ($T=2300$\,K) to the experimental value ($T=1823$\,K) than the prediction
of the BKS potential ($T=4800$\,K). For the properties of the glass, we find
that the CHIK potential predicts a structure that is in reasonably good
agreement with the one from the {\it ab initio} simulations, and quite
similar to the one of the BKS potential. In particular, it is remarkable
that the vibrational vDOS of the CHIK potential shows the double peak
feature at high frequencies, although this type of information has
not been entered at all in the optimization of the parameters. We also
mention that in Ref.~\cite{carre08} we have discussed the temperature
dependence of the diffusion constants and found that they are quite
similar to the one of the BKS potential~\cite{carre08}. In particular,
they show at intermediate and low temperatures an Arrhenius dependence
on temperature with an activation energy that is in good agreement with
the experimental values for silica. Finally, we mention that we have also
tested whether the CHIK potential is able to describe certain
properties of crystalline data and found that, {\it e.g.}, the
elastic constants are reproduced quite well~\cite{Carre_Phd}. Further
support that this potential is indeed quite reliable comes from
other recent simulations in which one has studied amorphous silica
at high pressures \cite{horbach08,zhang10}, its deformation at the
nanoscale \cite{zheng10}, as well as the modeling of silica nanopores
\cite{coasne11,cheng12,ori14,manga14,coasne14} and silica substrates
\cite{ewers12,lepinay15}. Although {\it a priori} one cannot expect
that the CHIK potential works well for these problems, since it has
been derived with respect to the properties of bulk amorphous silica,
in all these cases, the CHIK model has provided a reliable description.
Thus we can conclude that the underlying idea of the presented fitting
scheme works quite well in that the resulting potential is able to
describe with good accuracy properties of the system that have not been
used in the cost function.

Before we conclude we want to return to the surprisingly poor performance
of the approach proposed by Ercolessi and Adams. In Fig.~\ref{fig_force},
we have seen that the effective potential obtained by the force
matching method accurately reproduces the CPMD forces but that,
Fig.~\ref{gr_CPMD_FF_BKS}, the resulting local structure is very different
from the one predicted by the {\it ab initio} simulation. The reason
for this apparent discrepancy lies, most probably, in the fact that the
forces that act on the atoms of a local structure, say a tetrahedron, are
highly correlated with each other and that there is in fact a large amount
of compensations between these forces. The force matching method does not
take into account this effect, making that the local residual forces are
not necessarily small and thus in turn give rise to a local structure that
can be very different from the one of the target ({\it i.e.}~here the CPMD
structure). For the case of a metal, where the forces are not covalent
and much less directional, this effect can be expected to play a smaller
role, thus rationalizing why for such systems the force matching method
works significantly better than it is the case here~\cite{ercolessi94}.

In summary, we can conclude from the present work that using the local
structure as target for the optimization of the parameters of the
potential is a very promising approach. As it is set up, the method
gives a lot of freedom on the functional form that one wants to use,
and thus one can of course also use, {\it e.g.}, three body potentials. It is
also possible to include in the cost function additional quantities,
such as, {\it e.g.}, elastic constants etc. The future will tell how useful
the method will be for multi-component systems, but one can expect
that as long as the number of parameters does not become too large,
the approach will indeed allow to obtain reliable effective potentials.

\begin{acknowledgments}
We gratefully acknowledge financial support by Schott Glas and computing
time on the JUMP at the NIC J\"ulich, and on IBM/SP at CINES in
Montpellier. W.K. is member of the Institut Universitaire de France.

\end{acknowledgments}

\end{document}